\documentclass[a4paper,useams,usenatbib]{mnras}
\usepackage{graphicx}
\usepackage{subfig}
\usepackage{color}
\usepackage{xfrac}
\usepackage{xparse}
\usepackage{multirow}
\usepackage{multicol}
\usepackage{tabularx}
\usepackage{journals}
\usepackage[textwidth=1.2cm]{todonotes}
\usepackage{bm}
\usepackage{url}
\usepackage{times}
\usepackage[T1]{fontenc}    
\usepackage[utf8]{inputenc}
\usepackage[]{longtable}
\definecolor{darkgreen}{rgb}{0.0,0.5,0.0}
\definecolor{darkred}{rgb}{0.5,0.0,0.0}
\definecolor{brown}{rgb}{0.65,.16,0.16}
\definecolor{grey}{rgb}{0.4,0.5,0.6}

\newcommand{\deltam}{$\Delta \mathcal{M}_{12}$}
\newcommand{\Mabs}{$\mathcal{M}_{r}^{\rm Petro}$}
\newcommand{\mr}{$m_{r}^{\rm Petro}$}
\newcommand{\Mvir}{$\log(M_{\rm halo}/{\rm M_{\odot}})$}
\newcommand{\MvirNoUnit}{$\log M_{\rm halo}$}
\newcommand{\Mstar}{$\log(M_{\star} / {\rm M}_{\odot})$}
\newcommand{\MstarNoUnit}{$\log M_{\star}$}
\newcommand{\veldisp}{$\log \left[ \sigma_{{\rm e}8}/({\rm km~s}^{-1}) \right]$}
\newcommand{\veldispNoUnit}{$\log \sigma_{{\rm e}8}$}
\newcommand{\ageFifty}{$\log {\rm age}_{50}$}
\newcommand{\ageNinety}{$\log {\rm age}_{90}$}
\newcommand{\ageFiftyNinety}{$\log ({\rm age}_{50}/{\rm age}_{90})$}
\newcommand{\alphaFe}{$[\alpha/{\rm Fe}]$}
\newcommand{\mgb}{${\rm Mg}b$}
\newcommand{\fet}{${\rm Fe}3$}
\newcommand{\mgfeL}{$[{\rm Mg}/{\rm Fe}]'$}

\RequirePackage{color}\definecolor{RED}{rgb}{1,0,0}\definecolor{BLUE}{rgb}{0,0,0.6} 

\title[Stellar populations of BGGs versus magnitude gap]
{Do the stellar populations of the brightest two group galaxies depend on the magnitude gap?}
\author[M. Trevisan, G. A. Mamon and H. G. Khosroshahi]{M. Trevisan$^{1}$\thanks{E-mail: trevisan@iap.fr}, G. A. Mamon$^{1}$, H. G. Khosroshahi$^{1,2}$\\
 $^{1}$Institut d'Astrophysique de Paris (UMR 7095: CNRS \& UPMC, Sorbonne Universit\'es), 98 bis Bd Arago, 75014 Paris, France \\
 $^{2}$School of Astronomy, Institute for Research in Fundamental Sciences (IPM), P.O. Box 19395-5531, Tehran, Iran}
 
\date{Accepted 2016 October 4. Received 2016 September 21; in original form 2016 April 26; \textcolor{red}{Consolidated version with Erratum}}
\begin{document}

\label{firstpage}
\pagerange{\pageref{firstpage}--\pageref{lastpage}} \pubyear{2016}

\maketitle

\begin{abstract}
We investigate how the stellar populations of the inner regions of the first and the second
brightest group galaxies (respectively BGGs and SBGGs) vary as a function of
magnitude gap, using an SDSS-based sample of $550$ groups with
elliptical BGGs. The sample is complete in redshift, luminosity and for \deltam\ up
to 2.5 mag, and contains $59$ large-gap groups (LGGs, with \deltam\ $>
2.0$ mag). We determine ages, metallicities, and SFHs of BGGs and SBGGs using the {\sc STARLIGHT} code
with two different single stellar population models (which lead to
important disagreements in SFHs), and also compute \alphaFe\ from spectral indices.
After removing the dependence with galaxy velocity dispersion or with stellar mass, there is no correlation with
magnitude gap of BGG ages, metallicities, \alphaFe, and
SFHs. The lack of trends of BGG SFHs with magnitude gap suggests that BGGs in LGGs
have undergone more mergers than those in small-gap groups, but these
mergers are either dry or occurred at very high redshift, which in either case
would leave no detectable imprint in their spectra. 
We show that SBGGs in LGGs lie significantly closer to the BGGs (in projection) than
galaxies with similar stellar masses in normal groups, which appears to be a sign of the earlier
entry of the former into their groups. Nevertheless, the stellar population properties of the
SBGGs in LGGs are compatible with those of the general population of galaxies with similar
stellar masses residing in normal groups.
\end{abstract}


\section{Introduction}
\label{Sec_intro}
Galaxy groups are believed to play a key role in the formation
and evolution of galaxies.  There are several observational pieces of
evidence suggesting that the environment where galaxies reside affects their
evolution, changing their properties \citep{Oemler:1974, Dressler:1980,
  Weinmann+06, Peng.etal:2010, vonderLinden.etal:2010, Mahajan.etal:2011}. Determining how
environmental processes operate is of particular importance to understand the
formation of massive elliptical galaxies, which are known to reside
preferentially in the cores of groups. Gravitational phenomena such as interactions and mergers of
galaxies are more frequent in high-density environments than in the field
\citep{Mamon:1992}. Also, dynamical friction \citep{Chandrasekhar:1943a} dissipates orbital energy and angular momentum of the satellite galaxies, driving them to the group
centre until they finally merge with the central galaxy in a few Gyr
\citep{White:1976, Schneider.Gunn:1982, Mamon:1987b, Ponman.etal:1994}. Therefore, the
formation and evolution of the brightest galaxies in the Universe are
expected to be closely linked to the growth of their host groups.

The $r$-band absolute magnitude gap (hereafter \emph{magnitude gap} or simply
\emph{gap}), which we denote \deltam, 
between the group first and second-ranked galaxies,
in $r$-band luminosity (hereafter BGG for \emph{Brightest Group Galaxy} and
SBGG for \emph{Second Brightest Group Galaxy}),
is often used as a diagnostic of past mergers among
the more massive (or luminous) galaxies in groups
\citep{Ponman.etal:1994, vandenBosch.etal:2007, Dariush.etal:2010, Hearin.etal:2013}. 

Many studies have shown that there is a class of systems characterized by
bright isolated elliptical galaxies embedded in haloes with X-ray
luminosities comparable to those of an entire group. The magnitude gap
observed in these systems point to a very unusual luminosity function (LF), and these objects --
usually referred as \emph{fossil groups} (hereafter FGs) -- have been puzzling the
astronomical community for over two decades, since their discovery by
\cite{Ponman.etal:1994}. \cite{Jones.etal:2003} established the first formal
definition, in which a system is classified as fossil if their bolometric
X-ray luminosity is greater than $L_{X,{\rm bol}} \geq 10^{42}
h^{-2}_{50}$~erg~s$^{-1}$, and if they are large-gap groups (LGGs), i.e., the magnitude gap within $0.5$ virial radius is
\deltam~$> 2$~mag (in the $r$-band). 

The origin and nature of FGs is still matter of debate. It is not
clear if the merger scenario \citep{Carnevali.etal:1981, Mamon:1987b} can account for groups with such large magnitude
gaps, and alternative scenarios have been proposed to explain the formation
of these systems. They could be ``failed groups'' with atypical
LFs missing high-luminosity ($\approx L^*$) galaxies, while 
including very high ones
\citep{Mulchaey.Zabludoff:1999}. Based on differences between the isophotal
shapes of central galaxies located in FGs (which are often disky)
and in normal groups (which have boxy shapes), \cite*{Khosroshahi.etal:2006b}
suggested that, unlike the central galaxies of normal groups, the
first-ranked galaxies in FGs were produced in a wet merger event at
high redshift.

Studies of the evolution of FGs in cosmological simulations seem to suggest
that the mass assembly histories of FG haloes differ from those of small luminosity gap groups \citep{DOnghia.etal:2005, Dariush.etal:2007, Raouf.etal:2014}. 
A natural question is whether such a difference in the evolution of FGs is imprinted in the global properties of the stellar
content and in the star formation history of the brightest group galaxies. The detailed reconstruction of the 
stellar assembly of galaxies is, therefore, a powerful tool to constraint different scenarios and processes taking 
place during the formation and evolution of these objects.

It is still not clear whether the stellar population properties of FG BGGs differ from 
those of BGGs located in regular groups. Several studies show that FGs have lower optical to X-ray luminosity ratios compared to regular systems 
\citep{Jones.etal:2003, Yoshioka.etal:2004, Khosroshahi.etal:2007, Proctor.etal:2011}, indicating that FGs have abnormally high
$M_{\rm halo}/L_r$ at given $M_{\rm halo}$ (but see \citealp{Voevodkin.etal:2010}, \citealp{Harrison.etal:2012}, and \citealp{Girardi.etal:2014} for an alternative view). 
While these results are usually interpreted as a lack of cold baryons at given halo mass, one may wonder 
whether part of those trends is caused by differences in the stellar population properties. 
If it is the case, galaxy properties that affect the $M_{\star}/L$ ratios, such as age and metallicity, may vary with the magnitude gap, 
and the high $M_{\star}/L$ ratios of FG BGGs could indicate that these systems are older and/or more metal-rich than those residing in normal groups.

\citet{LaBarbera.etal:2009} used Sloan Digital Sky Survey Data release 4 (SDSS-DR4) and ROentgen SATellite (ROSAT) All Sky Survey X-ray data to define a sample
of $25$ FGs, and compared their properties with ``field\footnote{Their sample was not selected from a group catalogue. The magnitude gaps of the optical FG candidates 
and the control sample were defined within a cylinder
with radius of $350$~kpc and $\Delta z = 0.001$ centered on the galaxy.}''
galaxies selected from the same dataset.  
The examination of stellar populations revealed that FG BGGs have similar ages, metallicities, 
and $\alpha$-enhancements compared to the field galaxies. 
More recently, \citet{Eigenthaler.Zeilinger:2013} determined the age and metallicity gradients in a sample of six BGGs in FGs 
from deep long-slit observations with ISIS spectrograph at the \emph{William Herschel} Telescope. They found that  
metallicity gradients are weak ($\nabla {\rm [M/H]} = -0.19 \pm 0.08$), while age gradients are negligible 
($\nabla {\rm age} = 0.00 \pm 0.05$), suggesting that FGs are the result of multiple major mergers. The comparison of their results
with gradients in early-type galaxies determined by \citet{Koleva.etal:2011} indicates that FG BGGs follow similar 
$\nabla {\rm age}, \nabla {\rm [M/H]}$ vs. stellar mass trends as regular elliptical galaxies.
However, these studies suffer from low number statistics, therefore limiting their conclusions. 

Finally, previous studies show that, at fixed halo mass, the stellar mass of BGGs in FGs are greater than those of their
counterparts in regular groups \citep[e.g.][]{DiazGimenez.etal:2008,Harrison.etal:2012}. 
Since galaxy properties (e.g. morphology, colour, ages, and metallicities)
seem to be strongly related to stellar mass \citep{Balogh.etal:2009,
McGee.etal:2011, Trevisan.etal:2012, Woo.etal:2013}, the differences between fossil and regular groups -- if they exist --
could be at least partially related to differences in stellar mass.  

In this paper, we aim to obtain a clearer picture of the formation of the brightest galaxies and their host groups by 
studying how their stellar population properties and SFHs vary with the magnitude gap after correcting for the variations with the 
velocity dispersion, which is a proxy for the galaxy stellar mass. 
We use a large sample of $550$ SDSS groups at $0.015 \leq z_{\rm group} \leq 0.07$ and more massive than \Mvir~$\ge 13.1$.
Our sample is complete up to \deltam~$= 2.5$~mag, and contains $59$ groups with \deltam~$\ge 2.0$~mag.

This paper is organized as follows. In Section \ref{Sec_groups}, we describe
the sample of groups and the data used in our analysis. In Section \ref{Sec_m12}, we investigate how the properties of groups
and their brightest two galaxies vary with \deltam. In Section \ref{Sec_discussion}, we discuss the possible implications of our findings to the origin of FGs. Finally, in Section \ref{Sec_summary}, we present the summary and the conclusions of our study.
All masses and distances are given in physical units unless otherwise specified.
To be consistent with the group catalogue used in this study, 
we adopt, throughout this paper, a  $\Lambda$CDM cosmology with $\Omega_{\rm M} =
0.238$, $\Omega_{\Lambda} = 0.762$, $\Omega_{\rm b} = 0.042$ and 
$H_0 = 73$~km~s$^{-1}$~Mpc$^{-1}$.

\section{Data}
\label{Sec_groups}
\subsection{Sample selection}
\label{Sec_sample}

The galaxy groups were selected from the updated version of the catalogue
compiled by \cite{Yang.etal:2007}.\footnote{The group catalogue is publicly
  available at \url{gax.shao.ac.cn/data/Group.html}.}
The group catalogue contains 468,822 groups drawn from a sample of 593,617
galaxies from the Main Galaxy Sample \citep{Strauss.etal:2002} of the SDSS-Data Release 7 (DR7, \citealp{Abazajian.etal:2009}) 
database and 3234 galaxies with redshift determined by alternative surveys: Two Degree Field Galaxy
Redshift Survey (2dFGRS, \citealp{Colless.etal:2001}), Point Source Catalog
Redshift Survey (PSCz, \citealp{Saunders.etal:2000}), and
the Third Revised Catalog of Galaxies (RC3,
\citealp{deVaucouleurs.etal:1991}).\footnote{This corresponds to
  sample2\_L\_petro among the \citeauthor{Yang.etal:2007} group catalogues.}

The radii $r_{180}$, i.e, the radii of spheres that are $180$ times denser than the \emph{mean} density of the Universe,
are derived from the $M_{180}$ masses given in the \citeauthor{Yang.etal:2007}
catalogue, which are based on the ranking of Petrosian luminosities. 
We then calculated the virial radii ($r_{\rm vir} = r_{200}$, 
where $r_{200}$ are the radii of spheres that are $200$ times denser than the 
\emph{critical} density of the Universe) and 
masses ($M_{\rm halo} \equiv M_{\rm vir} = M_{200} = 100\,H^2(z)\,r_{200}^3/G$) by assuming the 
NFW \citep*{Navarro.etal:1996} profile and the concentration-mass relation given by 
\cite{Dutton.Maccio:2014}.\footnote{See appendix~\ref{Ap_mean_to_critical}
  for the conversion of Yang's virial radii to our definition.}

%
\begin{figure}
\centering
\includegraphics[width=\hsize]{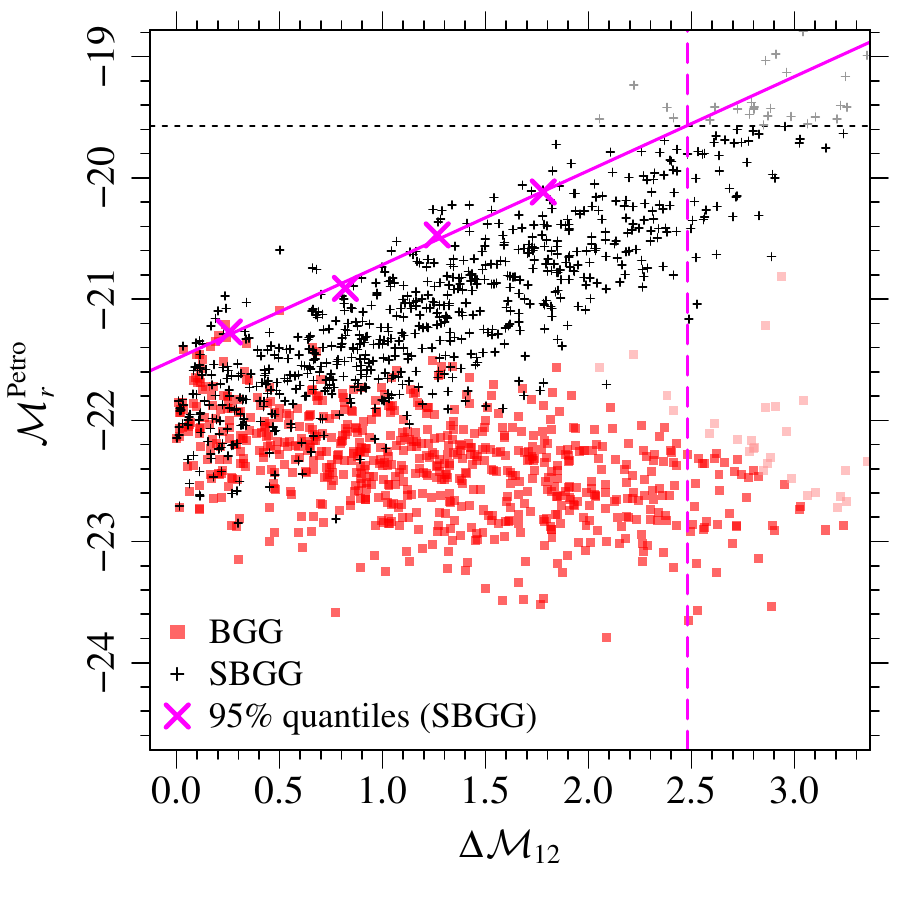}
\caption {Luminosity versus magnitude gap diagram to extract the
completeness limit for the magnitude gap of our sample. The BGGs and SBGGs are shown as \emph{red} and 
\emph{black symbols}. The \emph{solid magenta lines} indicates the linear fit to
the 95-percentiles of $\mathcal{M}_{r,2}^{\rm Petro}$ in bins of \deltam. 
The 95\% completeness magnitude limit of our full sample is shown as the \emph{black dashed horizontal line} (\Mabs~$=$~$-19.57$).
Therefore, the 90 percent complete magnitude gaps for the full sample is found where the solid magenta line meets the
dashed horizontal line.
}
\label{Fig_delta_compl}
\end{figure}

We selected groups that satisfy the following criteria:

\begin{enumerate} 
\item \label{sample_z} redshifts in the  range from $0.015$ to $0.07$;

\item \label{sample_richness} at least 5 member galaxies;

\item \label{sample_Mhalomin} masses \Mvir~$\geq 13.1$;

\item \label{sample_Lmin} at least two member galaxies within $0.5\,r_{\rm vir}$ brighter than
  \Mabs~$\leq -19.57$, where 
\Mabs\ is the k-corrected SDSS Petrosian absolute magnitude in the $r$ band; 
  
\item \label{sample_gap} the magnitude gap, defined as the difference between the k-corrected SDSS
$r$-band Petrosian absolute magnitudes of the BGG and SBGG
galaxies within half the  virial radius, i.e.,
\[
\Delta \mathcal{M}_{12} = \mathcal{M}_{r,2}^{\rm Petro} - \mathcal{M}_{r,1}^{\rm Petro} \ ,
\]
\noindent is smaller than $2.47$~mag;

\item \label{sample_zoo} first-ranked galaxy classified as an elliptical galaxy according to the
  information retrieved from Galaxy Zoo~1 project database
  \citep{Lintott.etal:2011};
  
\item \label{sample_pps} the relative velocity of the SBGG with respect to the BGG must be 
$< 2.7 \sigma_{\rm los} (R)$, where $\sigma_{\rm los} (R)$ is the predicted line-of-sight group
velocity dispersion at a distance $R$ from the BGG.
\end{enumerate}

%
\begin{figure}
\centering
\includegraphics[width=0.9\hsize]{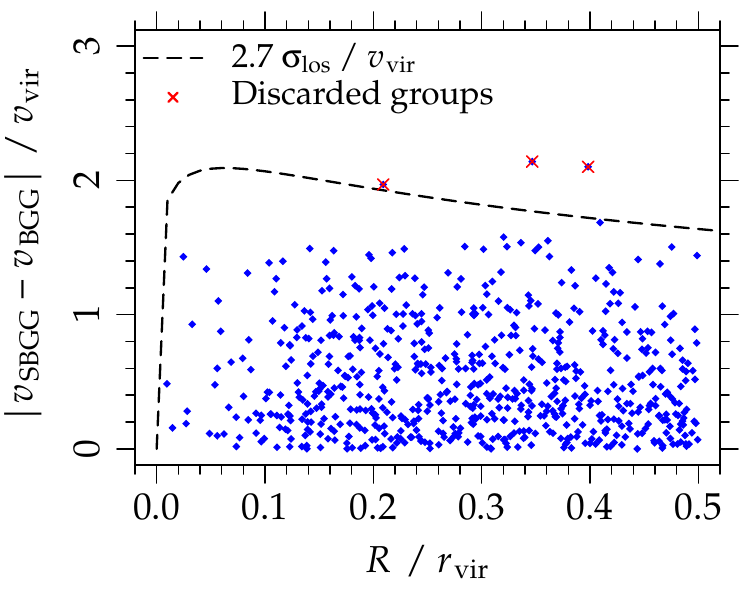}
\caption{Distribution of SBGGs in projected phase space (line-of-sight
  velocity relative to the BGG versus projected distance to the BGG). The
  velocities and distances are normalized to the virial velocity ($v_{\rm
    vir}$, see text) and radius ($r_{\rm vir}$), respectively. The \emph{dashed line}
  corresponds to $2.7 \sigma_{\rm los}(R)/v_{\rm vir}$, where $\sigma_{\rm los} (R)$ is the
  predicted line-of-sight velocity dispersion at a projected distance $R$ from the BGG (see text). The groups
  that lie above the curve are discarded from the sample (\emph{red crosses}).}
\label{Fig_vlos}
\end{figure}

The lower redshift limit was chosen to avoid selecting groups too close to the edge of the catalogue (the groups were defined using galaxies at $0.01 < z_{\rm gal} < 0.2$). 
The upper limit was optimized to obtain the largest possible number of groups with \deltam~$\ge 2$~mag, given the other criteria and taking into account the variation of \Mabs\ and \deltam\ limits with $z$.

The \MvirNoUnit, \Mabs, and \deltam\ completeness limits were established as follows. We compared the halo mass function of our sample with the theoretical halo mass function computed using the {\sc NumCosmo} package\footnote{\url{http://www.nongnu.org/numcosmo/}} \citep{NumCosmo:2014}. The adopted halo mass lower limit, \Mvir~$\geq 13.1$, corresponds to the
value above which the difference between the observed and theoretical mass functions is smaller than~$\sim 0.1$~dex. 

The sample limit in absolute magnitude cannot be simply derived from our adopted maximum redshift and the limit of extinction-corrected apparent magnitudes of the  SDSS
MGS (which would yield \Mabs~$=-19.64$), because the the reference frame for the k-corrections is at $z = 0.1$ and different galaxies have slightly different k-corrections. We therefore determine the 95 percent limit in  absolute magnitudes following 
the geometric approach similar to that described by \cite{Garilli+99} and \cite{LaBarbera.etal:2010a}. We first determine the 95 percentile of \mr\ in bins of \Mabs\ and then perform a linear fit to the 95-percentile points, so that the the value of  \Mabs\ where the best-fit line intersects \mr~$=~17.77$ defines the absolute magnitude
of 95 percent completeness. This leads to a 95 percent completeness limit of \Mabs~$\leq -19.57$ 
for our sample. 

This absolute magnitude limit in turns leads to a sample complete up to \deltam\ $= 2.47$ mag, as illustrated in Fig.~\ref{Fig_delta_compl}. 

Finally, the upper limit for the absolute value of the relative line-of-sight velocity between the
SBGG and the BGG was adopted to avoid projection effects, as illustrated in
Fig.~\ref{Fig_vlos} (where the virial velocity is defined 
with $v_{\rm vir}^2 \equiv v_{200}^2 = G\,M_{200}/r_{200}=10\,H(z)\,r_{200}$).
This predicted limit, $2.7 \sigma_{\rm los}$, was computed by assuming a single-component NFW profile of
concentration $c~=~6$ (the value expected for haloes with \Mvir~$\sim 13.4$),
with a velocity anisotropy that varies with physical radius according to the
formula of \cite{Mamon&Lokas05b}, which is a good approximation to the
measured velocity anisotropy in simulated $\Lambda$CDM haloes \citep{Mamon+10}.

The criteria above lead to a sample of 657  groups. 
The conclusions of this work depend little on the precise values used in
these selection criteria.
The k-corrections were obtained with the {\sc kcorrect} code (version 4\_2)
of \cite{Blanton.etal:2003a},
choosing as reference the median redshift of the SDSS MGS ($z = 0.1$).\footnote{Although the median redshift of our sample is 0.05, we kept the median redshift of the SDSS MGS as the reference for the k-corrections. The difference between $\mathcal{M}^{0.1}_{r, {\rm Petro}}$ and $\mathcal{M}^{0.05}_{r, {\rm Petro}}$ is $\sim 0.1$ mag, and has no effect on the results and conclusions of our study.}

\subsection{Spectroscopic incompleteness}

The completeness of the SDSS spectroscopy in high-density regions is affected by the fibre collision limit, which prevents neighboring fibres from being closer than $55''$. 
This spectroscopic incompleteness might affect the correct indentification of the BGGs and SBGGs.
To address this issue, we used the SDSS photometric catalogue to identify galaxies that could be BGGs or SBGGs, but have no SDSS-DR7 spectra. 

We first investigate if there are galaxies within $1~r_{\rm vir}$ from the luminosity-weighted center of each group that are brighter than the BGG of that group. 
We then used the SDSS-DR12 redshifts (when available\footnote{The spectra of many objects that are not in SDSS-DR7 are now available in DR12. However, these objects were observed with the BOSS (Baryon Oscillation Spectroscopic Survey) spectrograph, whose fibres are smaller than those of the SDSS spectrograph ($2''$ rather than $3''$). The difference in the fibre aperture might introduce bias in our results and, for this reason, we opted for not using these new spectra in our study. Therefore, we discard the groups with incomplete SDSS-DR7 spectroscopy.}) to check if the galaxy is within the redshift range of the group, i.e.,
\begin{equation}
 |z - z_{\rm group}|\ c < (2.7\ \sigma_{\rm group})\ ,
 \label{Eq_dz}
\end{equation}
If there is no SDSS-DR12 spectrum, we discard the group.
We discard $12$ groups that contain galaxies brighter than the BGG with redshifts
in the range given by equation~(\ref{Eq_dz}) and
$30$ groups with galaxies brighter than the BGG but no spectroscopic redshifts.

Following a similar approach, we retrieved from the photometric
catalogue all galaxies within $0.5~r_{\rm vir}$ from the BGGs that are
brighter than the SBGG of that group and have spectroscopic redshifts
according to Eq.~(\ref{Eq_dz}), or that do not have spectroscopic redshift
available ($31$ and $34$ objects, respectively). 
We discard all the $65$ groups that contain galaxies that follow
these criteria. 

In summary, following the criteria \ref{sample_z} to \ref{sample_pps} listed in Section~\ref{Sec_sample}
but discarding a total of $107$ groups with incomplete SDSS-DR7 spectroscopy we obtain our final sample of $550$ groups, among which $59$ have \deltam~$> 2$~mag.

\subsection{Galaxy properties}
\label{Sec_galprops}

The galaxy magnitudes, stellar masses, velocity dispersions, and specific
star formation rates (sSFR) were retrieved from the SDSS-DR12 
database\footnote{The stellar mass and sSFR estimates correspond to the parameters
{\tt lgm\_tot\_p50} and {\tt specsfr\_tot\_p50} from the SDSS table {\tt galSpecExtra} 
\citep{Kauffmann.etal:2003,Brinchmann.etal:2004b,Salim.etal:2007}.}
\citep{Alam.etal:2015}. 
The match between the Yang et al. catalogue and the SDSS-DR12 sample was performed with {\tt TOPCAT}\footnote{\url{http://www.star.bris.ac.uk/~mbt/topcat/}} \citep{topcat} by assuming a maximum
separation of $5''$ between the sky positions and a maximum difference in
redshift of $\Delta z < 0.0005$ (i.e. velocity differences less than $150 \, \rm km \, s^{-1}$). 
The stellar masses and sSFR correspond to the estimates available in the MPA-JHU spectroscopic catalogue, and described in \cite{Brinchmann.etal:2004b}.
We also retrieved the {\tt eClass} parameter from the SDSS-DR7 database, which is based on the first two eigencoefficients from the Principal Component Analysis (PCA) of galaxy spectra 
\citep{Yip.etal:2004}. The {\tt eClass} parameter ranges from about $-0.35$ to $0.5$ for early- to late-type galaxies.

The velocity dispersions are measured through the fixed aperture of the SDSS fibre (diameter of $3''$), therefore they need to be normalized to the same physical aperture. 
We assume that the velocity dispersion profile is well described by
$\sigma_{\rm ap} = \sigma_{{\rm e}8} \left[{R_{\rm ap}}/{(R_{\rm e} / 8)}
  \right]^{-0.066}$ \citep{Cappellari.etal:2006}, where $R_{\rm e}$ is the
effective radius of the galaxy (containing half the projected luminosity) and normalize the velocity
dispersions to $\sigma_{\rm e8}$, which corresponds to the dispersion
measured through an aperture with a radius of one eighth of the effective
radius.

\subsubsection{Star formation histories and metallicities}
\label{Sec_starlight}

We derived ages and metallicities from the SDSS spectra
using the {\sc STARLIGHT} spectral fitting code 
\citep{CidFernandes.etal:2005}. For each galaxy, {\sc STARLIGHT} combines
single stellar population (SSP) model spectra of given age and metallicity
and returns the contribution, 
as a percentage of mass, from each basis SSP. This distribution traces directly the SFH. For each galaxy
in the sample, we determine the ``cumulative'' mass fraction,
i.e., the fraction of stars older than a given age, as a function
of age. We then interpolate the cumulative SFH of each galaxy to obtain the
galaxies ages, defined as the age when half of the stellar mass was already
formed  
(i.e., age~$ \equiv {\rm age}_{50}$), and the age when $90$\% of the stellar mass was formed (${\rm age}_{90}$). 
From the {\sc STARLIGHT} results, we also compute the mass-weighted metallicities.

%
\begin{figure}
\centering
\includegraphics[width=\hsize]{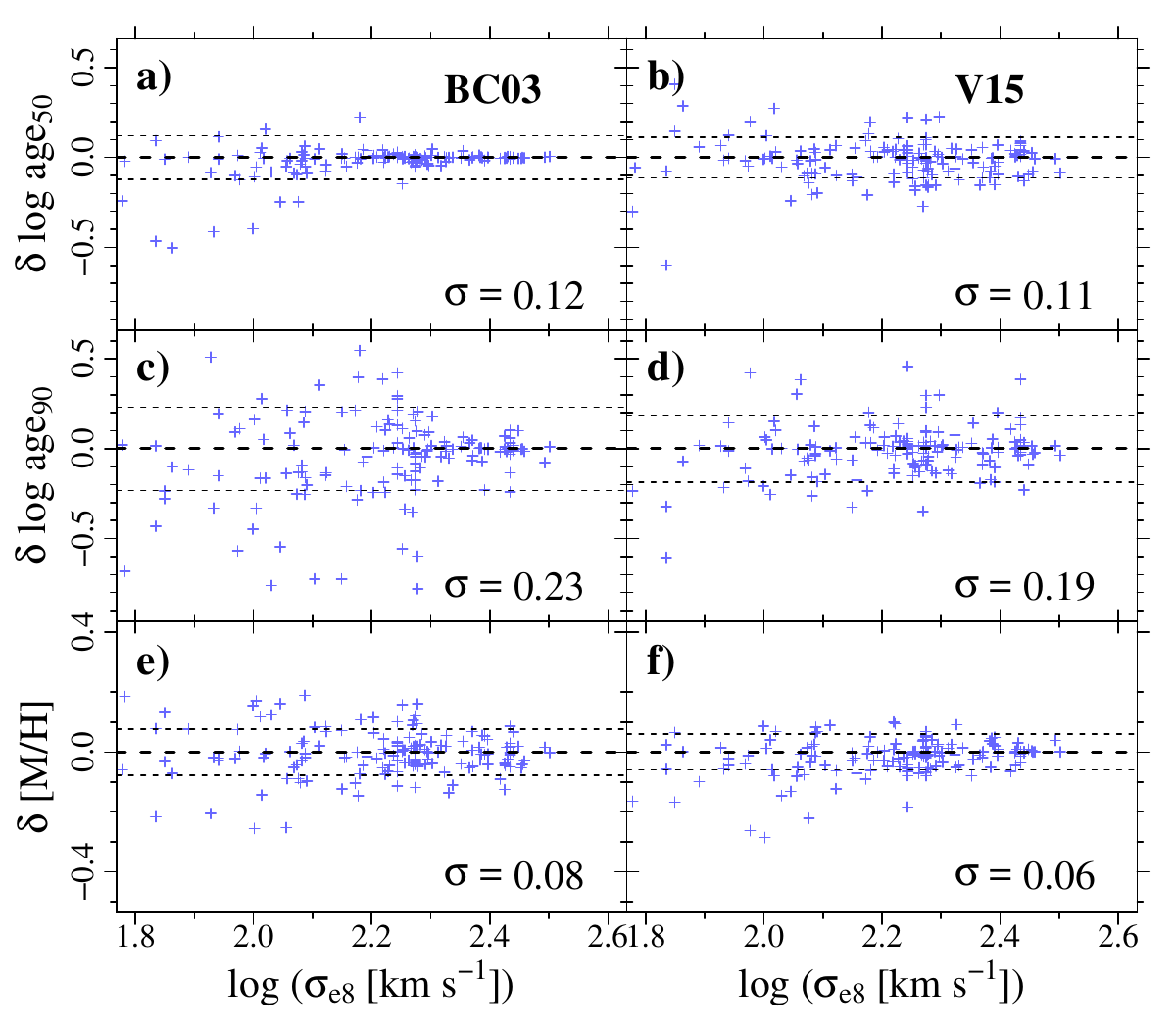}
\caption {Uncertainties in ages and metallicities, estimated from differences between the properties derived from multiple spectroscopic observations of the same objects ($\delta$). The \emph{left} and the \emph{right panels} show the results obtained with BC03 and V15 models, respectively. 
The \emph{long} and the \emph{short dashed lines} indicate
$\delta = 0 \pm \sigma$, where $\sigma$ is the standard deviation of the distribution of~$\delta$. 
}
\label{Fig_age_met_errors}
\end{figure}

To check the dependency of the stellar population properties with the models used, we selected two sets of SSP 
models. 
One of them is based on the Medium resolution Isaac Newton Telescope Library of Empirical Spectra \citep[MILES, ][]{SanchezBlazquez.etal:2006}, using the updated version 10.0 \citep[][hereafter V15]{Vazdekis.etal:2015} of the code presented in \cite{Vazdekis.etal:2010}. 
We selected models computed with \cite{Kroupa:2001} universal initial mass function (IMF), and isochrones from BaSTI 
(Bag of Stellar Tracks and Isochrones, \citealp{Pietrinferni.etal:2004, Pietrinferni.etal:2006}).
We also used \cite[][BC03]{Bruzual.Charlot:2003} models, calculated with Padova 1994 
evolutionary tracks \citep{Bressan.etal:1993,Fagotto.etal:1994a,Fagotto.etal:1994b,Girardi.etal:1996} and with the \cite{Chabrier:2003} IMF.
The BC03 model employs the STELIB stellar library \citep{LeBorgne.etal:2003}.
For both sets of models, we adopted a constant $\log {\rm age}$ step of $0.2$, and the grids cover ages from $0.03$ to $13.5$~Gyr. 
For the BC03 models, we selected SSPs with metallicities [M/H]~=~$\{-1.7, -0.7, -0.4, 0.00, +0.40\}$ and the grid of V15 models contains SSPs with
[M/H]~=~$\{-1.26, -0.66, -0.25, +0.06, +0.15, +0.40\}$.

The fit was performed in the wavelength region from $3800$ to $7400$~\AA.
Before running the code, the observed
spectra are corrected for foreground extinction and de-redshifted, and
the SSP models are degraded to match the wavelength-dependent resolution
of the SDSS spectra, as described in
\cite{LaBarbera.etal:2010a} and \cite{Trevisan.etal:2012}. We adopted the \cite*{Cardelli.etal:1989} extinction
law, assuming $R_{\rm V} = 3.1$.

To estimate the uncertainties in the stellar population properties, we retrieved from the SDSS database all objects in our sample that have multiple spectroscopic observations. We found 168 galaxies, 62 of them being elliptical, with multiple spectra with signal-to-noise (S/N) ratio greater than 5. 

In Fig.~\ref{Fig_age_met_errors}, we show the differences between the properties derived from multiple spectroscopic observations as a function of velocity dispersion for galaxies of all morphological types. 
After dividing the values of $\sigma$ shown in Fig.~\ref{Fig_age_met_errors} by $\sqrt{2}$, 
we find that the typical errors in \ageFifty, \ageNinety, and ${\rm [M/H]}$ 
are $0.08$, $0.16$, and $0.06$ (BC03), and $0.08$, $0.13$, and $0.04$ (V15).
If we consider only elliptical galaxies, whose spectra tend to have higher
S/N ratios, the typical uncertainties (in dex) are $1.2$ to $3$ times smaller:

\[
\begin{array}{lll}
 \sigma({\rm age}_{50}, {\rm age}_{90}, [{\rm M/H}])_{\rm E}
& \!\!\!= (0.03,\, 0.10,\, 0.04)  & {\rm (BC03)} \ , \nonumber \\
                                                            
& \!\!\!= (0.06,\, 0.08,\, 0.04)  & {\rm  (V15)} \ .\nonumber
\end{array}
\]


\subsubsection{$[\alpha/{\rm Fe}]$ ratios}
\label{Sec_alpha}

The $\alpha$-elements (${\rm O}$, ${\rm Mg}$, ${\rm Si}$, ${\rm S}$, ${\rm Ca}$, ${\rm Ti}$) are synthesized primarily in supernovae (SN) type II 
(e.g. Arnett 1978, Woosley \& Weaver 1995), while SNs Ia yield mostly iron-peak elements with little $\alpha$-element production (e.g. Nomoto et al 1984, Thielmann et al 1986). 
Since SN Ia events are delayed relative to SNs II\footnote{For typical elliptical galaxies, the peak of SN Ia rates occur $\sim 0.1 - 1$~Gyr later than that of SN II rates \citep[see e.g.][]{Matteucci.Tornambe:1987,Thomas.etal:1999,Matteucci.Recchi:2001}.}, the \alphaFe\ ratios are believed to be closely related to the star-formation 
timescales of galaxies \citep{Tinsley:1979, Thomas.etal:2005, delaRosa.etal:2011, Walcher.etal:2015}. 

To estimate the \alphaFe\ ratios of our BGGs and elliptical SBGGs, 
we adopted the approach described in \citet{LaBarbera.etal:2013} and \citet{Vazdekis.etal:2015}, which is based on the spectral indices \mgb\ and ${\rm Fe}3$\footnote{${\rm Fe}3 = ({\rm Fe}4383 + {\rm Fe}5270 + {\rm Fe}5335) / 3$ \citep{Kuntschner:2000}}. 
We measured the line-strengths with the {\sc Indexf}\footnote{\url{http://pendientedemigracion.ucm.es/info/Astrof/ellipt/pages/page4.html}} code \citep{Cardiel:2010}, and 
applied corrections for the broadening due to the internal velocity dispersion of the galaxy following the prescriptions of \citet{delaRosa.etal:2007}.

%
\begin{figure}
\centering
\includegraphics[width=0.9\hsize]{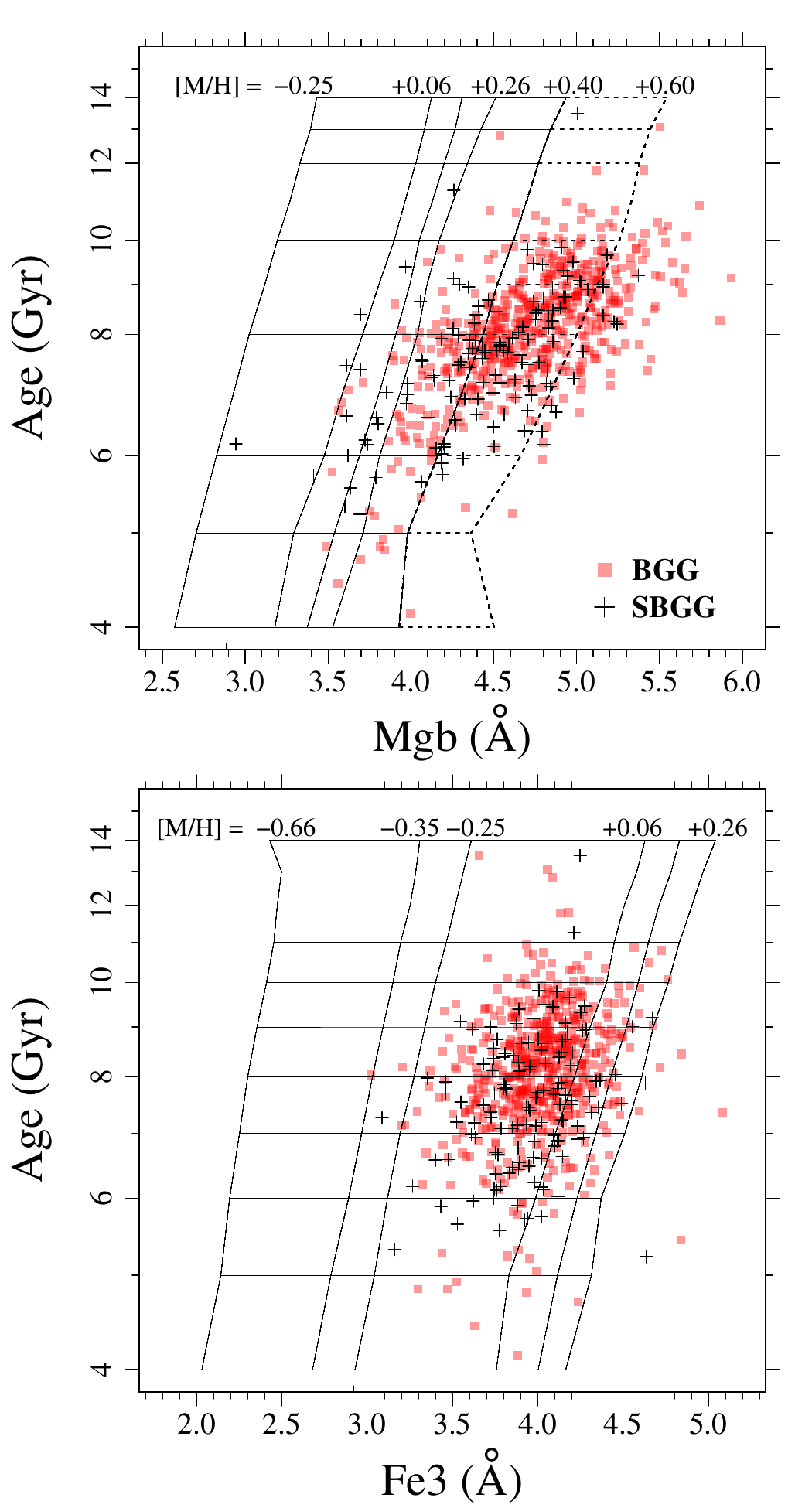}
\caption {Illustration of the procedure to obtain the \alphaFe\ proxy. The luminosity-weighted ages obtained using {\sc STARLIGHT} with V15 models are shown
as a function of the spectral indices \mgb\ (\emph{upper panel}) and ${\rm Fe}3$ (\emph{lower panel}). The BGGs and the SBGGs are represented by the \emph{red} and the 
\emph{black} symbols, respectively, and the V15 model grids are indicated in both panels. 
The dashed part of the grid in the upper panel indicates our linear
extrapolation of the model \mgb\ to $[{\rm M}/{\rm H}] = +0.6$. 
}
\label{Fig_ind_grid}
\end{figure}

The procedure to determine the proxy of \alphaFe\ is illustrated in Fig.~\ref{Fig_ind_grid}, where we show the BGG and the SBGG luminosity-weighted ages (derived using {\sc STARLIGHT} with V15 models, see Section~\ref{Sec_starlight}) as a function of \mgb\ and ${\rm Fe}3$, as well as the predictions from the V15 models with different metallicities.
For each galaxy, we estimate two independent metallicities, $Z_{{\rm Mg}b}$ and $Z_{{\rm Fe}3}$, by fixing the galaxy age and interpolating the model grid. 
As discussed by \cite{LaBarbera.etal:2013}, estimating $Z_{{\rm Mg}b}$ of an $\alpha$-enhanced population may require extrapolation of the models to higher metallicities. 
This is illustrated in the upper panel of Fig.~\ref{Fig_ind_grid}, where we show our linear extrapolation of the model \mgb\ to metallicity $[{\rm M}/{\rm H}] = +0.6$.
To reduce the uncertainties in the interpolated and extrapolated values, the model grids include all metallicities $\geq -0.66$ available for the V15 models, i.e., 
[M/H]~=~$\{-0.66, -0.35, -0.25, +0.06, +0.15, +0.26, +0.40\}$.
Finally, the proxy of \alphaFe\ is then defined as the difference between these two metallicities, $[Z_{{\rm Mg}b} / Z_{{\rm Fe}3}] = Z_{{\rm Mg}b} - Z_{{\rm Fe}3}$.

\begin{figure}
\centering
\includegraphics[width=0.9\hsize]{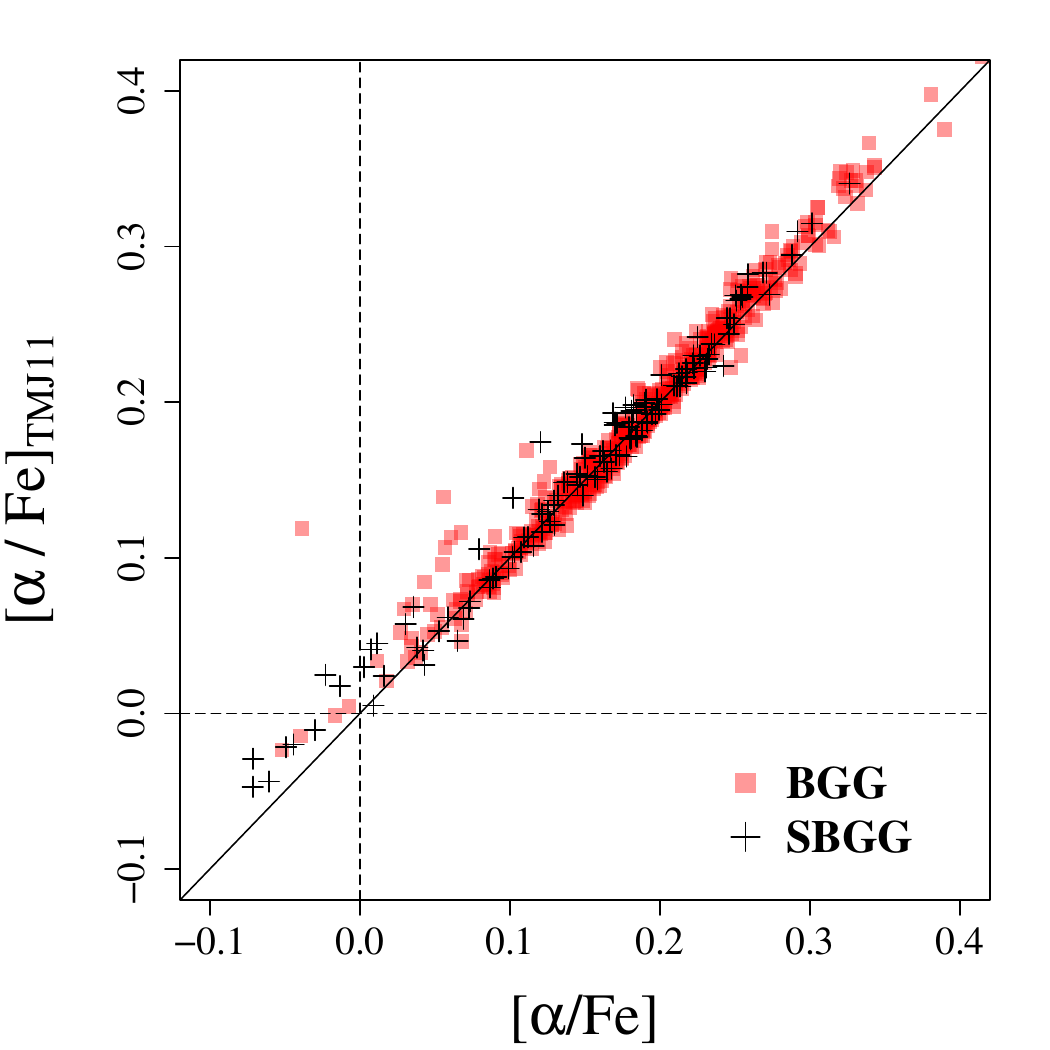}
\caption {The \alphaFe\ ratios estimated with \citet[][TMJ11]{Thomas.etal:2011} models as a function of \alphaFe\ obtained 
with the V15 models. The latter corresponds to the proxy  $[Z_{{\rm Mg}b} / Z_{{\rm Fe}3}]$ 
calibrated according to $[\alpha/{\rm Fe}] = -0.07 + 0.51\ [Z_{{\rm Mg}b} / Z_{{\rm Fe}3}]$ (Eq.~\ref{Eq_alpha}).
}
\label{Fig_alpha}
\end{figure}

We also computed \alphaFe\ by taking the
\alphaFe\ ratios explicitly into account. To this event, we compared the  
\mgfeL\footnote{\mgfeL~$ =\sqrt{{\rm Mg}b\, (0.72\, {\rm Fe}5270 + 0.28\, {\rm Fe}5335)}$ \citep{Thomas.etal:2003}}, 
\mgb, and \fet\ indices to the predictions by 
\citet[][hereafter TMJ11]{Thomas.etal:2011}. 
First, we estimate the metallicity [M/H]$_{\rm TMJ}$ from the \mgfeL\ indice, which is independent of \alphaFe\ (\citealp{Thomas.etal:2003}; see also \citealp{Vazdekis.etal:2015}), by fixing the galaxy
age and interpolating the TMJ11 model grid (i.e., we applied same procedure as illustrated in Fig.~\ref{Fig_ind_grid}, but using \mgfeL\ and different models).
We then fitted the models to obtain the predicted (polynomial) relation $[\alpha/{\rm Fe}] = \widetilde{p}({\rm Mg}b, {\rm Fe}3, [{\rm M/H}], {\rm age})$, 
and used this relation with {\sc STARLIGHT} ages and [M/H]$_{\rm TMJ}$ to obtain the $[\alpha/{\rm Fe}]_{\rm TMJ11}$ estimates.
As in the previous studies by \citet{LaBarbera.etal:2013} and \citet{Vazdekis.etal:2015}, we find a very tight correlation between $[\alpha/{\rm Fe}]_{\rm TMJ11}$ and
$[Z_{{\rm Mg}b} / Z_{{\rm Fe}3}]$. Selecting all elliptical galaxies within $0.5~r_{\rm vir}$ in our sample to fit the relation between these two quantities, we find
\begin{equation}
 [\alpha/{\rm Fe}]_{\rm TMJ11} = -0.07 + 0.51\ [Z_{{\rm Mg}b} / Z_{{\rm Fe}3}]\ , 
 \label{Eq_alpha}
\end{equation}
with a very small scatter of $0.019$~dex. If we consider only the BGGs and elliptical SBGGs, we obtain an even smaller scatter of $0.017$~dex.
Finally, in Fig.~\ref{Fig_alpha} we compare $[\alpha/{\rm Fe}]_{\rm TMJ11}$ with our final values of \alphaFe, i.e., the proxy $[Z_{{\rm Mg}b} / Z_{{\rm Fe}3}]$ calibrated according to Eq.~\ref{Eq_alpha}.

%

%
\begin{figure*}
\centering
\includegraphics[width=0.8\hsize]{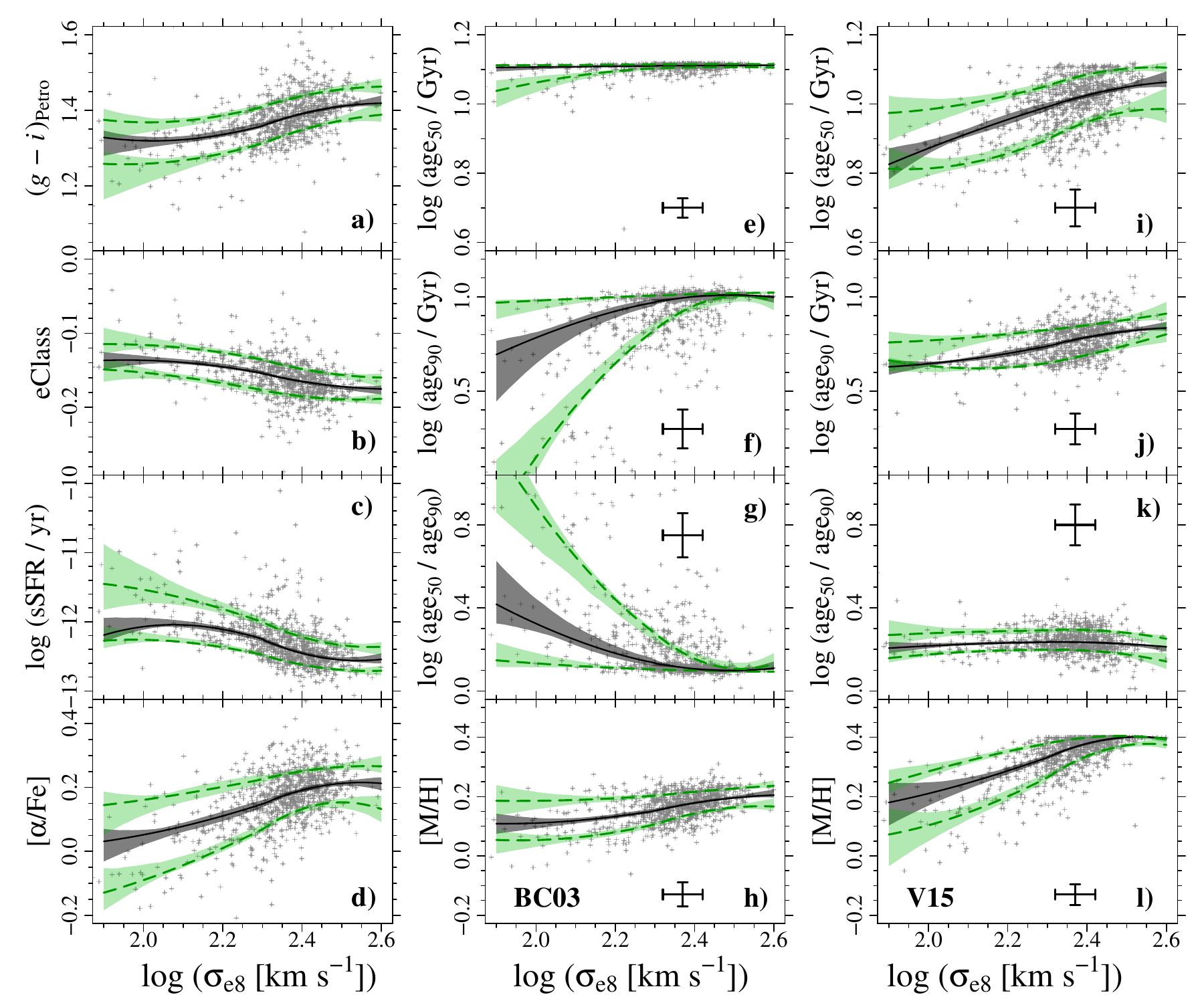}
\caption {Galaxy properties as a function of velocity dispersion normalized to an aperture with one-eighth of the effective radius. The \emph{left panels} ({\bf a--d}) 
present the extinction- and k-corrected $g-i$ colour, the parameter {\tt eClass}, the specific star formation rate, and \alphaFe\ as a function of the \veldispNoUnit.
The \emph{middle} and \emph{right panels} show the median ages (\ageFifty), the lookback times at which $90\%$ of the galaxy stellar mass was formed (\ageNinety), 
the star-formation timescales given by \ageFiftyNinety, and metallicities obtained with BC03 (panels {\bf e--h}) and V15 ({\bf i--l}) models.
All the elliptical galaxies within $0.5~r_{\rm vir}$ of our sample of groups are shown (\emph{black symbols}). Typical errors in ages and metallicities are indicated in panels {\bf e} to {\bf l} (see Section~\ref{Sec_starlight}).
The \emph{black solid} and the \emph{green lines} indicate the median, the $16$, and the $84$ percentiles of the galaxy properties in bins of velocity dispersion. The uncertainties in the relations are indicated as \emph{shaded areas} ($95\%$ confidence interval).
}
\label{Fig_prop_fit}
\end{figure*}

\subsection{Stellar population properties vs. velocity dispersion relations}
\label{Sec_fit_relations}

Many studies suggest that the properties of elliptical galaxies are more correlated to velocity dispersions than to stellar masses \citep[e.g.][]{Bernardi.etal:2003b, LaBarbera.etal:2014}. 
In addition, the uncertainties on \veldispNoUnit, as listed in the SDSS database, 
are typically $0.1$~dex, while the errors on stellar masses are $0.2$~dex \citep[e.g.][]{Duarte.Mamon:2015}.
Therefore, we analyse how the properties of the BGGs and the elliptical SBGGs vary with \deltam\ (see Section~\ref{Sec_m12_StarPop}) 
after correcting the relations with \veldispNoUnit\ instead of stellar masses. 

To fit the relations between galaxy properties and \veldispNoUnit, we selected all elliptical galaxies within $0.5~r_{\rm vir}$ in our 
sample groups and divided them in bins of \veldispNoUnit, with $\sim\!100$ galaxies per bin.
We adopt a non-parametric approach by computing the median values of the distributions of the galaxy properties in each bin and fitting these values using 
the locally-weighted polynomial regression method (LOESS). 
The relations were fit using the R function {\tt loess}\footnote{\url{https://stat.ethz.ch/R-manual/R-patched/library/stats/html/loess.html}} \citep{R:2015}. 
The fit is perfomed locally using the data points in the neighbourhood of $x$, 
weighted by their distance from $x$. The neighbourhood includes a proportion $\gamma$ of all the points; we adopted $\gamma = 1$, i.e., all points.
A 2nd-order polynomial is then fit using weighted least squares, and each data point $i$ receives the weight proportional to 
$\left( 1 - \left[d_i/{\rm max}(\vec{\bf d}) \right]^3 \right)^3$, where $d_i$ is the distance of the data point $i$ to $x$ and 
${\rm max}(\vec{\bf d})$ is the maximum distance among the data points in the neighbourhood of $x$. 

The fitted relations are shown in Fig.~\ref{Fig_prop_fit}. Some properties such as ages and sSFR show a large scatter at low velocity dispersions. 
To estimate how the scatter varies with \veldispNoUnit, we also computed the $16^{\rm th}$ and the $84^{\rm th}$ percentiles, 
$q_{16}$ and $q_{84}$, of the distributions in bins of velocity dispersion and fitted the percentiles versus \veldispNoUnit\ relations using the same approach described above.

For each galaxy, we then computed the normalized distance to the best fit:
\begin{equation}
 \delta P = \frac{P - f_{P}(\log \sigma_{{\rm e}8})}{{\rm IPR}_{P}(\log \sigma_{{\rm e}8})}
 \ ,
 \label{Eq_delta}
\end{equation}
where $f_{P}(\log \sigma_{{\rm e}8})$ is the best fit to the relation between the median value of 
the galaxy property $P$ and \veldispNoUnit, and ${\rm IPR}_P$ is the ``interpercentile''
range of the distribution of the property $P$ for galaxies with a given velocity dispersion, defined as:

\begin{eqnarray} 
 {\rm IPR}_P(\log \sigma_{{\rm e}8}) =   f_P(\log \sigma_{{\rm e}8}) - q_{16}(\log \sigma_{{\rm e}8})\ , &\mbox{}& \delta P < 0  \nonumber \\
 {\rm IPR}_P(\log \sigma_{{\rm e}8}) = q_{84}(\log \sigma_{{\rm e}8}) -    f_P(\log \sigma_{{\rm e}8})\ , &\mbox{}& \delta P > 0  \nonumber 
\end{eqnarray}

The errors in the fitted relations were estimated by bootstrapping the full sample $1000$ times, 
and the $95\%$ confidence intervals are shown as shaded areas in 
Fig.~\ref{Fig_prop_fit}.

\subsection{Differences in the star formation histories according to the
  single stellar population model}

The star formation histories of the BC03 and V15 models turn
out to be strikingly different for elliptical galaxies, as seen in
Fig.~\ref{Fig_prop_fit}: in comparison with the
BC03 model, elliptical galaxies (within half the virial radius of their
groups) with the V15 model have formed half of their stellar masses at lower redshifts.
While with BC03, elliptical galaxies were formed very quickly, with $50\%$ of their
total stellar masses already formed $~12$~Gyr ago (Fig.~\ref{Fig_prop_fit}e),
with V15, $50\%$ of the stellar mass is formed from $\sim 8$ to $11$~Gyr
ago, depending on their velocity dispersion (Fig.~\ref{Fig_prop_fit}i). 
These differences are also clearly seen in \ageNinety\
(compare Figs.~\ref{Fig_prop_fit}f and \ref{Fig_prop_fit}j) and in the
durations of star formation 
(compare Figs.~\ref{Fig_prop_fit}g and \ref{Fig_prop_fit}k).
In addition, according to V15 models, galaxies with \veldisp~$\sim 2.4$ have 
metallicities that are $\sim 0.2$~dex higher than the values obtained with BC03 models
(Figs.~\ref{Fig_prop_fit}h,l).
  
\section{Results}
\label{Sec_m12}

\subsection{Stellar masses and velocity dispersions of BGGs and SBGGs in groups with large and small gaps}
\label{Sec_mass_sigma}

\begin{figure}
\centering
\includegraphics[width=\hsize]{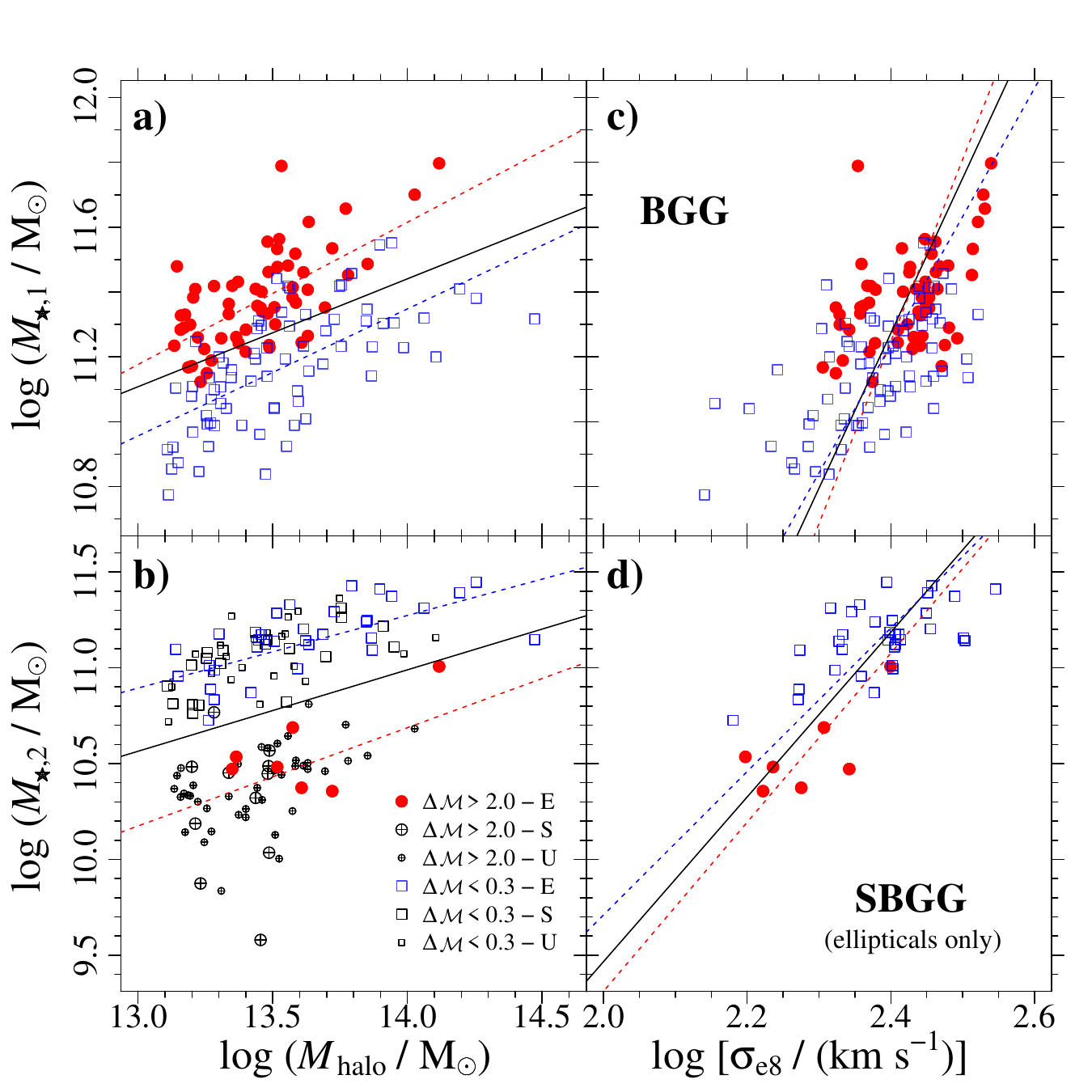}
\caption {Stellar mass as a function of halo mass ({\it left panels}) and velocity dispersion ({\it right panels})
  for the BGGs ({\it upper panels}) and SBGGs {\it lower panels} of groups
  with large (\deltam~$> 2.0$) and small (\deltam~$< 0.3$) magnitude gaps
  (respectively \emph{red dots} and \emph{blue open squares}).
  In panel ({\bf b}), we indicate the Galaxy Zoo morphologies of the SBGGs by different symbols: 
  elliptical (E), spiral (S), and galaxies with uncertain classification (U). 
  The \emph{black solid lines} shown in all panels are the best linear fits to the 
  \MstarNoUnit\ $vs.$ \MvirNoUnit\ (panels {\bf a,b}) and \MstarNoUnit\ $vs.$ \veldispNoUnit\ ({\bf c,d}) relations for all BGGs ({\bf a,c}) or SBGGs ({\bf b,d}) in our sample. 
  The \emph{red} and \emph{blue dashed lines} are the best fits for galaxies in groups with
  \deltam~$>2.0$ and $<0.3$, respectively. 
}
\label{Fig_fossils_E}
\end{figure}

In Fig.~\ref{Fig_fossils_E}, we compare the stellar masses and velocity dispersions of BGGs and SBGGs residing in groups 
with large ($> 2$~mag, $59$ groups) and small ($<0.3$~mag, 74 groups) magnitude gaps. 
For simplicity, we refer to groups with \deltam~$> 2$ and $<0.3$ as 
\emph{large-gap} (LGG) and \emph{small-gap groups} (SGGs), respectively. 
At fixed halo mass, the stellar masses of BGGs in LGGs are $\sim 0.2$~dex greater than
those of their counterparts in groups with small gaps (Fig.~\ref{Fig_fossils_E}a), as previously noted by \citet{DiazGimenez.etal:2008} and \citet{Harrison.etal:2012}. 
By construction, SBGGs in large-gap and small-gap groups have different \MstarNoUnit\ distribution; the difference in
stellar masses is $\sim 0.6$~dex in haloes with \Mvir~$\sim 13.5$ and $\sim
0.8$~dex for \Mvir~$\sim 13.1$ (Fig.~\ref{Fig_fossils_E}b). 
Nevertheless, the \MstarNoUnit\ versus \MvirNoUnit\ relations of BGGs and SBBGs in large-gap groups have slopes similar to that of their counterparts in SGGs. 
We fitted the $M_{\star} \propto M_{\rm halo}^{\alpha}$ relation and found 
$\alpha  =  0.44 \pm 0.06$ and $0.39 \pm 0.05$ for BGGs in LGGs and SGGs, respectively. 
For the SBGGs, we get $\alpha  =  0.51 \pm 0.11$ in LGGs and $\alpha  =  0.38 \pm 0.05$ in SGGs.

On the other hand, BGGs in large-gap and in small-gap groups follow slightly different $M_\star-\sigma_{{\rm e}8}$ relations, as
presented in Fig.~\ref{Fig_fossils_E}c. 
Assuming that $M_{\star} \propto \sigma_{{\rm e}8}^{\beta}$, we obtain $\beta = 4.78 \pm 0.06$ for all BGGs in our sample, 
$\beta = 3.95 \pm 0.16$ for BGGs in SGGs, and a slightly steeper relation with $\beta = 5.61 \pm 0.26$  for BGGs in large-gap groups.
Finally, we find that the slope of the SBGG $M_\star-\sigma_{{\rm e}8}$ relation is smaller than that of the BGG relation ($\beta = 4.30 \pm 0.08$, Fig.~\ref{Fig_fossils_E}d).

%

\subsection{Stellar populations and star formation histories versus magnitude gap}
\label{Sec_m12_StarPop}

%
\begin{figure*}
\centering
\includegraphics[width=\hsize]{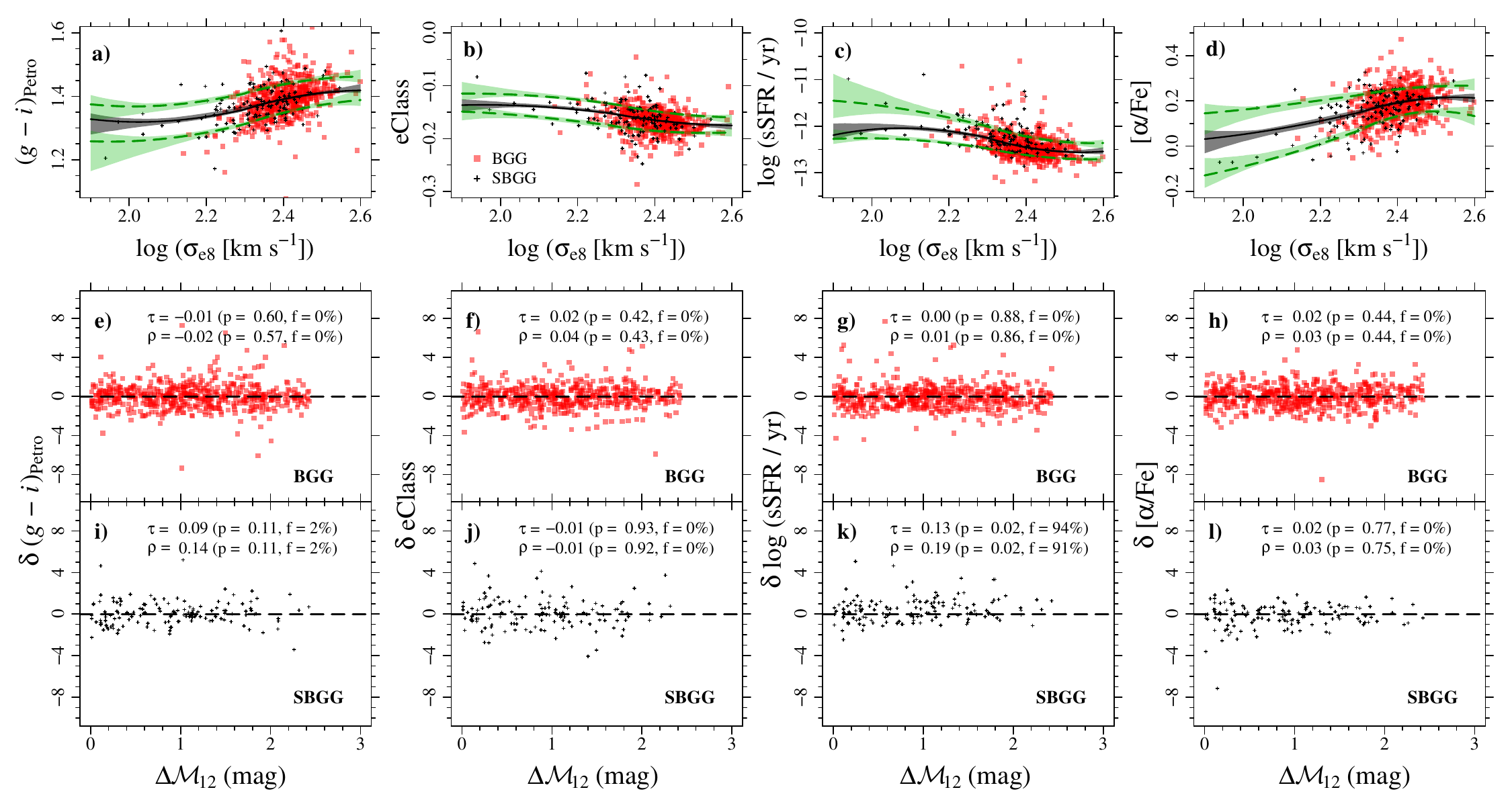}
\caption {Galaxy properties as a function of the magnitude gap. The upper
  panels ({\bf a--d}) present the extinction- and k-corrected $g-i$ colour, the parameter {\tt eClass}, 
  the specific star formation rate, and \alphaFe\ as a function of the \veldispNoUnit. 
  The first and the second-ranked galaxies are represented by the red and black symbols,
  respectively. The \emph{solid lines} indicate the best fit to the
  properties vs. velocity dispersion relations from Fig.~\ref{Fig_prop_fit}, and the \emph{green dashed lines} correspond the best fit
  to the $16^{\rm th}$ and $84^{\rm th}$ percentiles of the distribution in bins of \veldispNoUnit. 
  The uncertainties in the fitted relations are shown as the \emph{grey} and \emph{green shaded areas} (see Sect.~\ref{Sec_fit_relations}).
  The lower panels show the difference between first
  ({\bf e--h}) and second-ranked ({\bf i--l}) galaxy properties and the
  corresponding best-fit relation, normalized by the difference between the median and the $16^{\rm th}$ percentile (if $\delta < 0$) or the $84^{\rm th}$ 
  percentile and the median (if $\delta > 0$, Equation \ref{Eq_delta}). 
  In each panel, we indicate the Kendall ($\tau$) and Spearman
  ($\rho$) rank correlation coefficients, the $p$-values, and the fraction ($f$) of the $1000$ realizations of 
  the fit (see Section \ref{Sec_fit_relations}) that leads to $p < 0.05$. 
} 
\label{Fig_dColor_dMag_sigma}
\end{figure*}

%
\begin{figure*}
\centering
\includegraphics[width=\hsize]{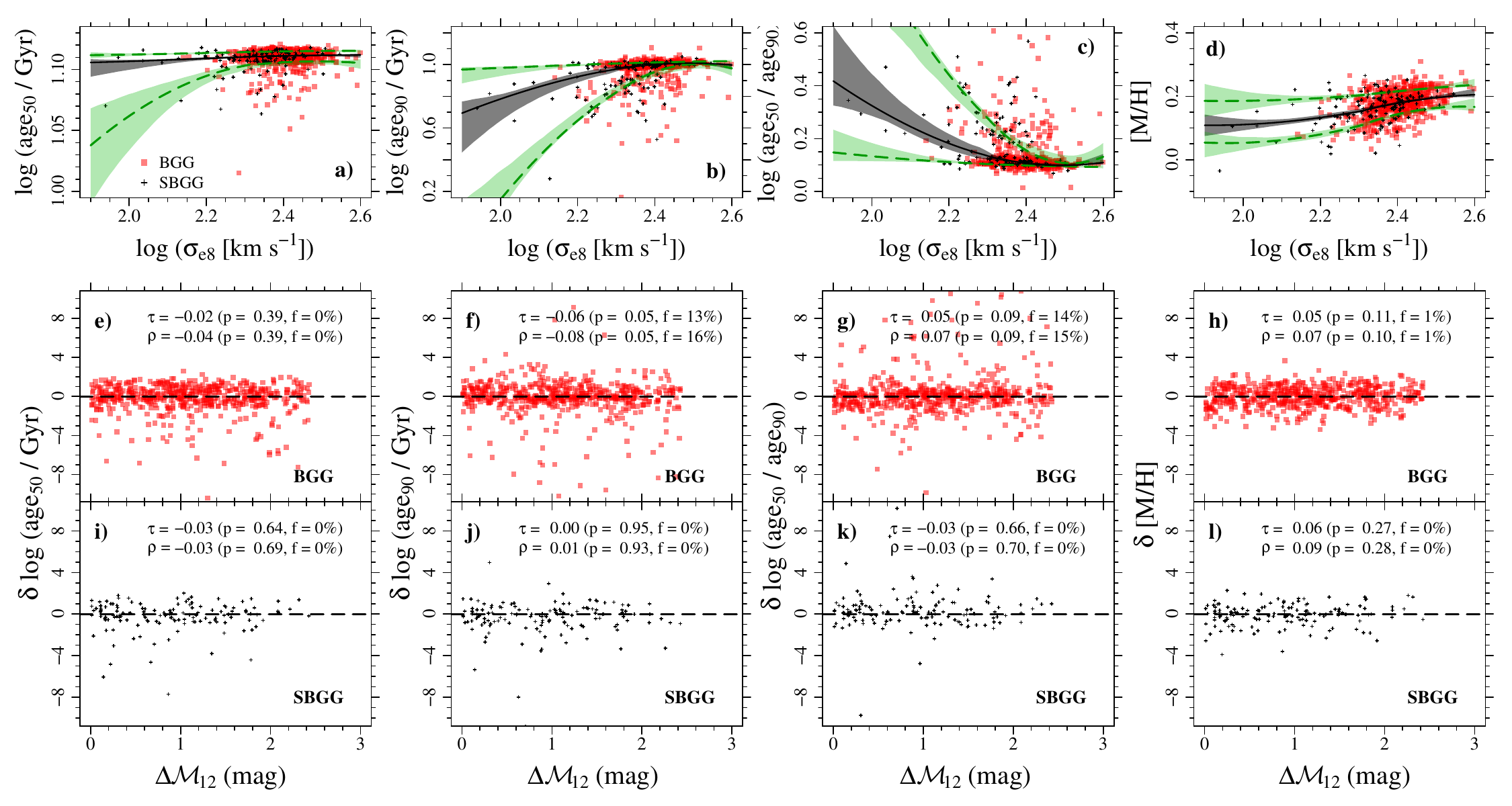}
\caption {Same as Fig.~\ref{Fig_dColor_dMag_sigma} (with the same notation), but for stellar ages and
  metallicities obtained with the BC03 model. The upper panels ({\bf a--d}) present \ageFifty, 
  \ageNinety, \ageFiftyNinety, and metallicities as a function of
  \veldispNoUnit. 
}
\label{Fig_dAgeMet_dMag_bc03_sigma}
\end{figure*}

%
\begin{figure*}
\centering
\includegraphics[width=\hsize]{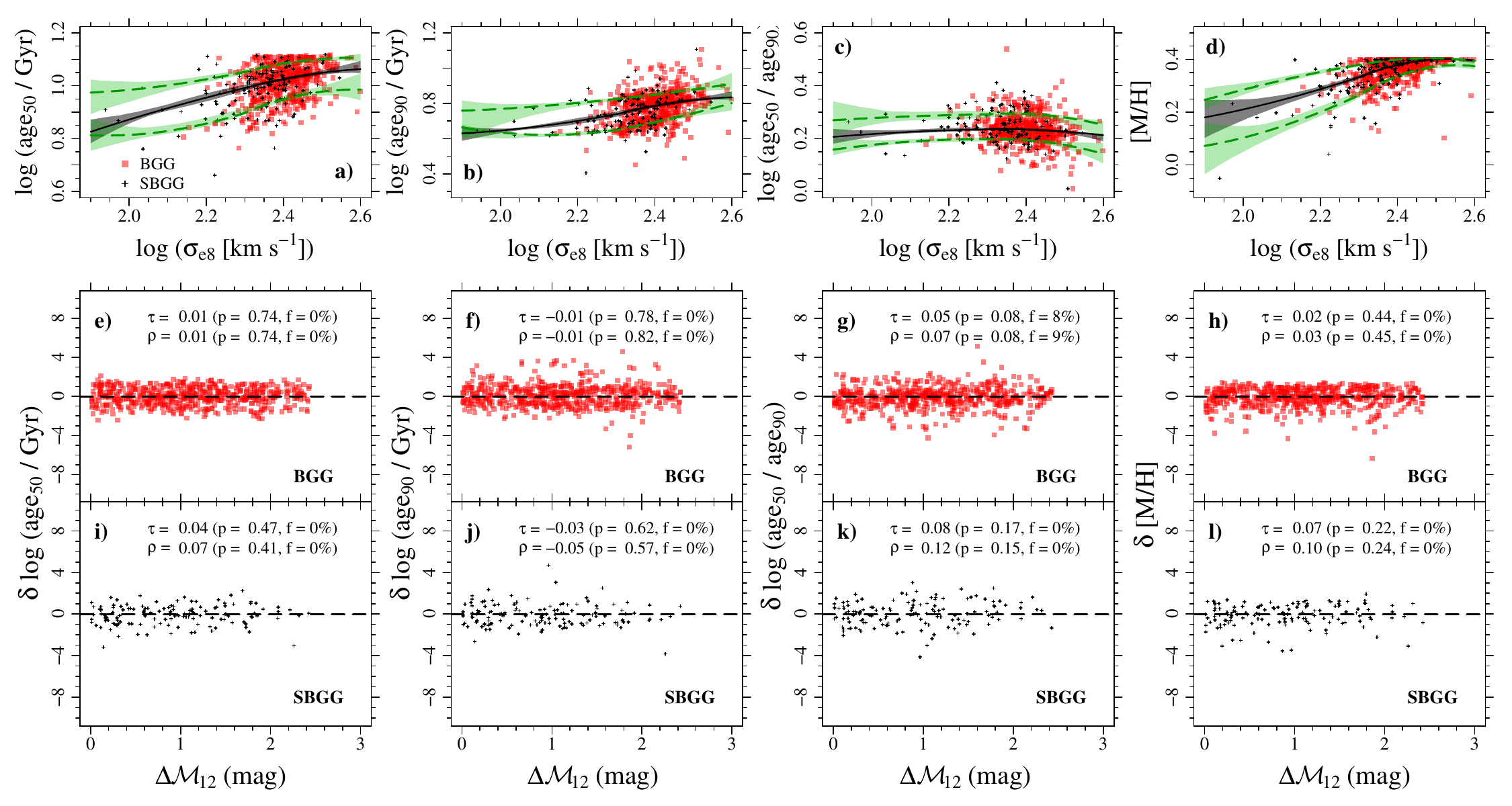}
\caption {Same as Fig.~\ref{Fig_dColor_dMag_sigma}, but for stellar ages and
  metallicities obtained with the V15 model. 
}
\label{Fig_dAgeMet_dMag_miles_sigma}
\end{figure*}

To explore the variations of the stellar population properties of elliptical BGGs and SBGGs with \deltam,
we first correct the dependency of these properties with the galaxy velocity dispersion, as described in Section~\ref{Sec_fit_relations}.
In Fig.~\ref{Fig_dColor_dMag_sigma}, we present the extinction- and
k-corrected galaxy $g-i$ colour, the {\tt eClass} parameter, the sSFR, and \alphaFe\ values of the BGGs and elliptical SBGGs as a function
of \veldispNoUnit\ (panels a-d). The figure also shows (in panels e--h for the
BGGs and i--l for the SBGGs) the residuals $\delta P$
(eq.~[\ref{Eq_delta}]) of these properties as a function of magnitude gap.
In each panel, we show the Kendall and Spearman correlation coefficients. We repeated the same procedure for stellar ages, metallicities, \ageNinety, and \ageFiftyNinety, as shown in Figs.~\ref{Fig_dAgeMet_dMag_bc03_sigma} (BC03) and \ref{Fig_dAgeMet_dMag_miles_sigma} (V15). 

Once we correct for the trend with velocity dispersion, the BGGs and SBGGs appear to share the same properties regardless of the magnitude gap of the group where they reside. 
The statistical tests indicate that there is no significant trend of $\sigma_{{\rm e}8}$-corrected properties with \deltam\ for all the galaxy properties analysed (Figs.~\ref{Fig_dColor_dMag_sigma}~to~\ref{Fig_dAgeMet_dMag_miles_sigma}, panels e--l), except for the positive trend of SBGG sSFRs with gap 
($\tau = 0.13$, $\rho = 0.19$, $p = 0.02$, Fig.~\ref{Fig_dColor_dMag_sigma}k). 
However, this trend appear to be a consequence of galaxies having different fractions of light within the SDSS fibre, as we discuss in the next Section. 

%
\begin{figure}
\centering
\includegraphics[width=0.9\hsize]{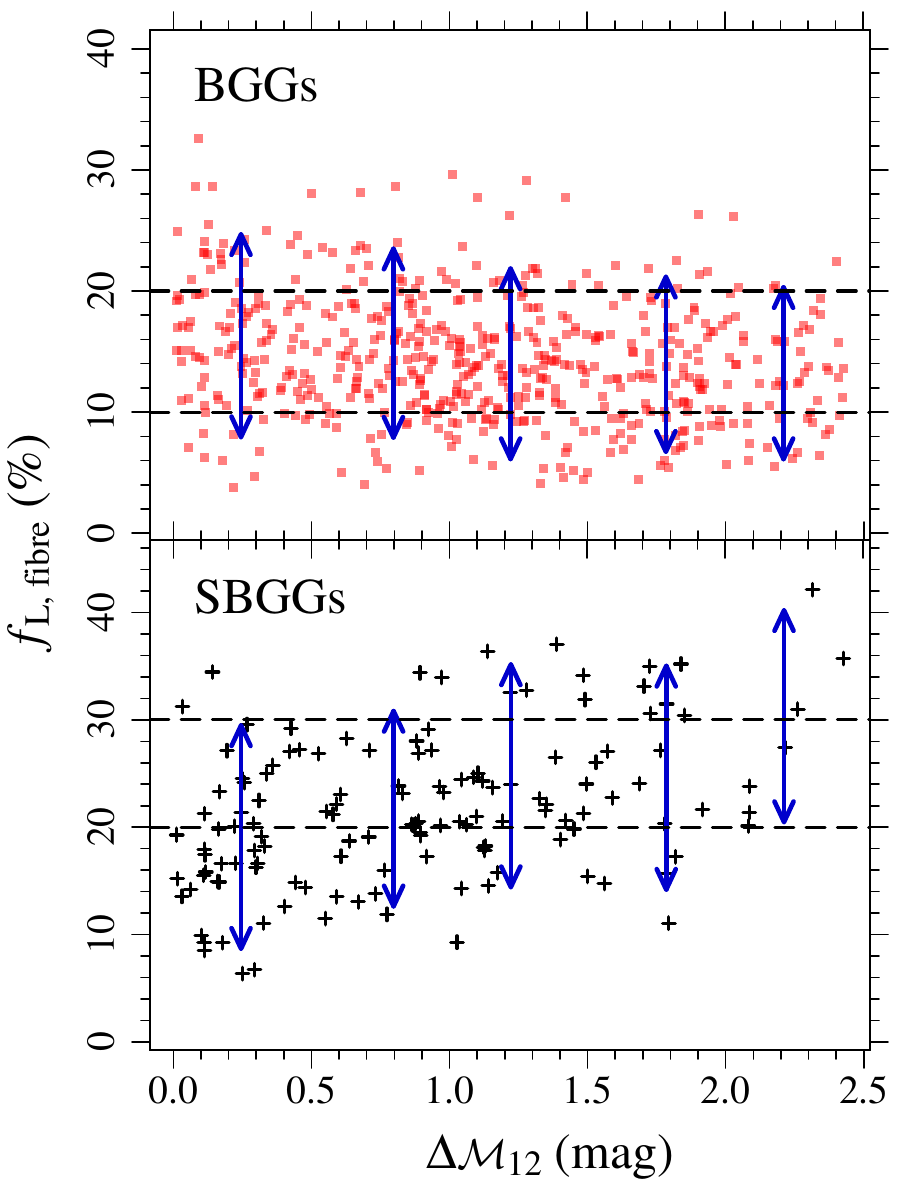}
\caption {Fraction of Petrosian $r$-band light within the SDSS fibre as a function of magnitude gap. The BGGs and the elliptical SBGGs are shown in the \emph{upper} and
\emph{lower panels}, respectively. The \emph{blue arrows} indicate the $5^{\rm th}$ and the $95^{\rm th}$ percentiles of the 
$f_{\rm L, fibre}$ distribution in bins of \deltam. The \emph{dashed lines} indicate the $f_{\rm L, fibre}$ limits of the subsamples defined to investigate 
the aperture effects (see Section~\ref{Sec_Aperture}).} 
\label{Fig_Lfrac_dMag}
\end{figure}

\subsubsection{Aperture effects}
\label{Sec_Aperture}

All the quantities derived from SDSS spectra are affected by the small aperture (diameter $= 3''$) of the fibre, which encompass only the inner part of the observed galaxy.
As discussed by several authors \citep[e.g.][]{Brinchmann.etal:2004b, Salim.etal:2007, Guidi.etal:2015}, the presence of age and metallicity gradients 
\citep{Koleva.etal:2011, Pilkington.etal:2012, LaBarbera.etal:2012, Eigenthaler.Zeilinger:2013, Hirschmann.etal:2015, Sanchez.etal:2015} can lead to biases and large uncertainties in the physical properties derived from the spectra. Although bright elliptical galaxies appear to have relatively flat age,
metallicity, and colour gradients \citep{Koleva.etal:2011, LaBarbera.etal:2012, Eigenthaler.Zeilinger:2013}, we investigate how our results are affected by aperture effects.

We computed the fraction of Petrosian $r$-band light within the fibre as $f_{\rm L} = {\rm dex} \left[-0.4 (m_r^{\rm fibre} - m_r^{\rm Petro}) \right]$, where $m_r^{\rm fibre}$ and $m_r^{\rm Petro}$ are the fibre and Petrosian magnitudes in $r$-band. The maximum value of $f_{\rm L}$ among the BGGs is $33\%$, and $99\%$ of our BGG sample have less than $28\%$ of the light within the fibre. On the other hand, the SDSS spectra of SBGGs contains a larger fraction of the galaxy total luminosity, with values of $f_L$ varying from $6$ up to $42\%$ (for elliptical SBGGs).
We, therefore, investigate if our results are affected by aperture effects by repeating the analysis for galaxies with similar values of $f_{\rm L}$. 
Since the range of $f_{\rm L}$ among the BGGs and the SBBGs is very different, we define different ranges of $f_{\rm L}$ to select the subsamples of BGGs and SBGGs, 
as we describe below.

In Fig.~\ref{Fig_Lfrac_dMag} we show how the fraction of light within the fibre varies as a function of \deltam. 
We indicate the $5^{\rm th}$ and $95^{\rm th}$ percentiles of the distribution of $f_{\rm L}$ in bins of gap for the BGGs and the SBGGs. 
For the BGGs, $f_{\rm L}$ decreases with increasing \deltam, and 95 percent of the BGGs in LGGs have $f_{\rm L} < 0.20$. 
On the other hand, the $5^{\rm th}$ ($95^{\rm th}$) percentiles of the SBGG $f_{\rm L}$ distributions in bins of \deltam\ increases from $0.08$ ($0.30$) for SBGGs in groups with \deltam~$< 0.3$ to $0.20$ ($0.41$) in LGGs. 

Therefore, to avoid selection effects when defining the subsamples of BGGs and SBGGs with similar $f_{\rm L}$, we require that the $f_{\rm L}$ limits are within the $5^{\rm th}$ and $95^{\rm th}$ of the distribution of $f_{\rm L}$ in bins of \deltam, as illustrated in Fig.~\ref{Fig_Lfrac_dMag}. 
We then define a subsample of $368$ BGGs with $10\% < f_{\rm L} < 20\%$ and another subsample of $63$ elliptical SBGGs with $20\% < f_{\rm L} < 30\%$, as shown in Fig.~\ref{Fig_Lfrac_dMag}. 

Following the approach presented in Section~\ref{Sec_fit_relations}, we fitted the galaxy properties vs. velocity dispersion relations of all elliptical galaxies that lie within $0.5$~r$_{\rm vir}$ from the BGG and have $10\% < f_{\rm L} < 20\%$ ($463$ objects). We repeated the procedure for a sample with $20\% < f_{\rm L} < 30\%$ ($220$ galaxies). The normalized distance to the best fit is computed using Eq.~(\ref{Eq_delta}), and the results are presented in Tables~\ref{Tab_tests_bgg} (BGGs) and \ref{Tab_tests_sbgg} (SBGGs).

The correlation coefficients obtained for the BGGs with similar $f_{\rm L}$ are very similar to the ones obtained for the whole sample, 
with no statistically significant trend with gap of any of the properties. On the other hand, the positive trend of sSFR with gap obtained for the whole sample of elliptical SBGGs (Fig.~\ref{Fig_dColor_dMag_sigma}k) is no longer observed when we restrict the analysis to SBGGs with similar values of $f_{\rm L}$, indicating that this trend might be due to aperture effects. 

\subsection{Second-ranked galaxies in large-gap groups}
\label{Sec_PSM}

The distribution of masses, magnitudes, and morphological types of SBGGs in large-gap and in small-gap
groups are very different, which makes the comparison between these galaxies
very difficult.  The fraction of elliptical galaxies among SBGGs in LGGs is smaller
than in SGGs ($12\%$ and $45\%$, respectively), with Barnard's test indicating high statistical significance ($p < 10^{-4}$). 
In Section \ref{Sec_m12_StarPop}, we focused on the variations of the elliptical SBGG properties with 
\Mabs, since galaxy properties versus stellar mass relations of other morphological types have large scatter and 
are not as well defined as those of elliptical galaxies.
Therefore, correcting the trends of properties with \veldispNoUnit\ (or \MstarNoUnit) for galaxies other than ellipticals is not straightforward.  

To overcome this issue and address the question whether LGG SBGGs of other morphological types have any peculiar property, we built
a control sample of galaxies with similar stellar masses and magnitudes as the general population of 
galaxies residing in groups with \deltam~$< 2.0$~mag by applying the Propensity Score Matching (PSM) technique (\citealp{PSM:1983}).  We
used the  {\sc MatchIt} package \citep{MatchIt:2011} written in 
R\footnote{\url{https://cran.r-project.org/}} \citep{R:2015}. This
technique allows us to select from the sample of galaxies
in groups with \deltam~$< 2.0$~mag a control sample in which the distribution of observed
properties is as similar as possible to that of the SBGGs in LGGs.  We adopted the logistic
regression approach \citep[see e.g.][]{Hilbe:2009, deSouza.etal:2015} to compute the propensity scores and the nearest-neighbour
method to perform the matching (see details in appendix~\ref{Ap_PSM}).

As we discuss in appendix~\ref{Ap_PSM}, LGGs are typically less massive than SGGs. 
We therefore limit our comparison to SBGGs residing in groups with \Mvir~$< 13.7$. 
The control sample was selected among all satellite galaxies (i.e., rank~$\ge
2$) within $0.5$~$r_{\rm vir}$, with $v_{\rm los} < 2.7 \sigma_{\rm v}(R)$ (see Fig.~\ref{Fig_vlos}), in groups with \deltam~$\leq 2.0$~mag and $13.1 \leq $~\Mvir~$\leq 13.7$.

\begin{figure}
\centering
\includegraphics[width=\hsize]{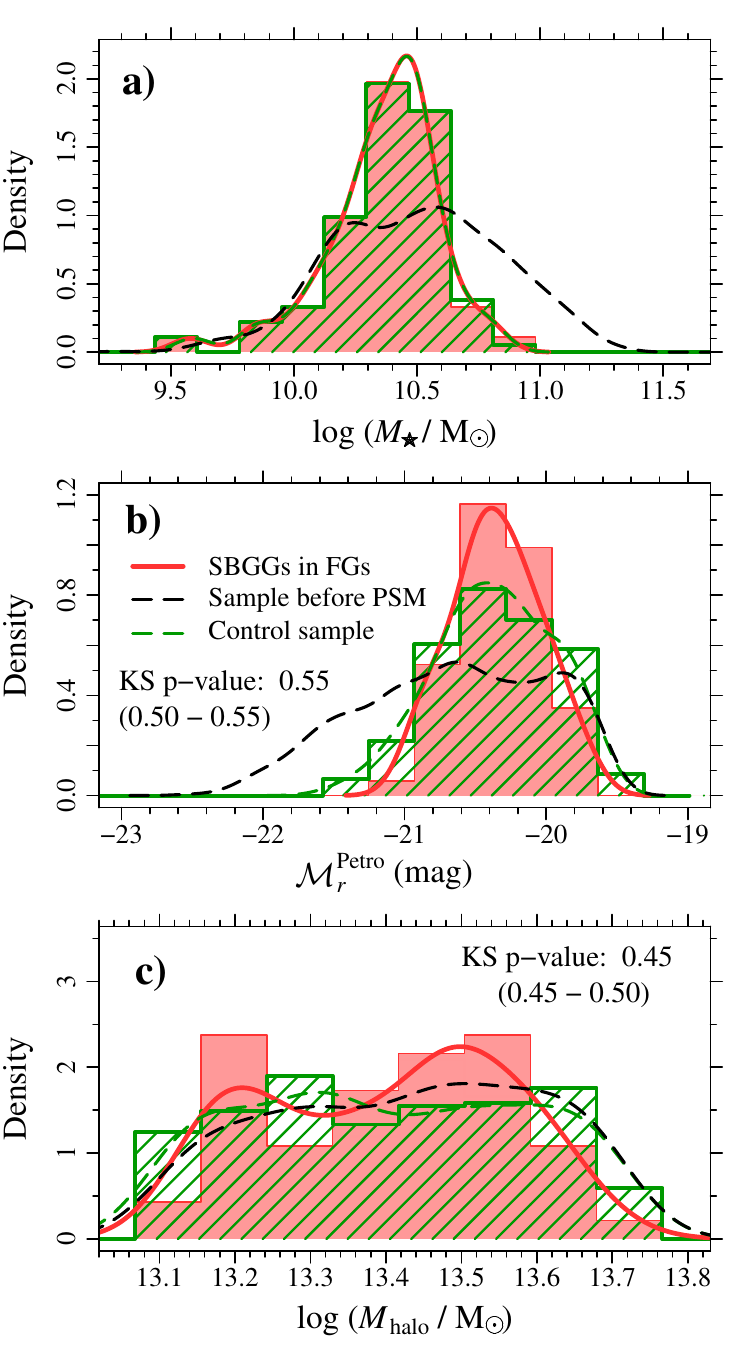}
\caption {Stellar masses ({\it top}), absolute magnitudes ({\it middle}), and
  halo masses ({\it bottom}) before and after the PSM (see text for
  details). The red histograms show the distributions
  of \MstarNoUnit, \Mabs, and \MvirNoUnit~of second-ranked galaxies in groups with
  \deltam~$> 2.0$~mag. The properties of the general population of galaxies in groups with \deltam~$<
  2$~mag (excluding the BGGs) are indicated by the \emph{black
  dashed lines}. The control sample, drawn from the
  sample of galaxies in groups with \deltam~$< 2$~mag by applying the PSM
  technique, is shown as the \emph{green histograms}.
The curves are obtained by smoothing the positions of the data points (not
the histograms) using a Gaussian kernel with standard deviation equal to one third
of the standard deviation of the data points. 
In panels {\bf b} and {\bf c} we indicate the $p$-value of the KS test (median, $1$ and $99$-percentiles).
}
\label{Fig_PSM}
\end{figure}

Figure \ref{Fig_PSM} shows the distribution of absolute magnitudes, stellar
masses and halo masses of the control sample before and after the PSM. We
performed the matching by stellar mass\footnote{Note that the PSM technique is not necessary when only one variable is used; 
other simpler matching methods would work as well as PSM. However, PSM allows us to test how the results change when taking other galaxy properties into 
account for the matching (see Table~\ref{Tab_PSM}).}, and Fig.~\ref{Fig_PSM}a illustrates the similarity of the distributions of stellar mass of the SBGGs and the control samples, as expected by construction. The distributions of SBGG absolute magnitudes and halo masses are compatible with those of the control sample, as shown in Fig.~\ref{Fig_PSM}b,c. 
The sample of SBGGs in LGGs have a smaller fraction of spirals ($20.8\%$ of the LGG SBGGs and $25.5\%$ of the control sample), and more ellipticals ($9.4$ and $8.4$ percent)
 and galaxies with uncertain morphological classification ($69.8$ and $65.1$ percent) than the control sample. 

As discussed in Section~\ref{Sec_Aperture} and shown in Fig.~\ref{Fig_Lfrac_dMag}, the fixed aperture of the SDSS fibre might introduce biases in our results. 
However, the $f_{\rm L}$ distributions of the LGG SBGGs and of the control sample are very similar, as indicated by a KS test ($p = 0.7$), and also by 
differences between their median values ($\Delta = 2.05\%$) and the quantile-quantile distances (${\rm Q-Q} = 1.5\%$, see Table~\ref{Tab_PSM}).  
 
In Figure \ref{Fig_rank2_control}, we compare the properties of the SBGGs in
LGGs with those of the control sample. We applied the KS test as a
comparison diagnostic, and we indicate the resulting $p$-value in each panel.
No clear distinction between the distributions of galaxy properties can be observed, with all $p$-values being greater than $0.1$.

In Fig.~\ref{Fig_Rrvir_sbgg}a, we revisit the projected phase space diagram
of SBGGs (see Fig.~\ref{Fig_vlos}) to distinguish LGGs from the control
sample. While the distribution of normalized line-of-sight velocities of the
SBGGs in LGGs resembles the corresponding distribution for the control sample,
there appears to be an excess of LGGs whose SBGGs lie closer to the BGG than
$\sim 0.25\,r_{\rm vir}$.
This appears more evident in Fig.~\ref{Fig_Rrvir_sbgg}b, which shows that the
radial distributions (in virial units) of LGG SBGGs is shifted to smaller distances compared to that 
of galaxies in the control sample. 
A KS test indicates that the shift of the distribution of SBGG $R/r_{\rm vir}$ to smaller radii in LGGs is significant ($p=0.02$).
The presence of a dominant BGG tends to
bring the SBGGs closer in, which should lead to short times for the next
merger between the BGG and the SBGG when the gap is large.

We repeated the matching by adding galaxy morphology and absolute magnitude in addition to stellar
mass. As can be seen in Table~\ref{Tab_PSM}, adding morphology and magnitude leads to similar results, i.e., there are no differences between the 
distribution of colour, ages, and metallicities of the SBGGs in LGGs and
the control sample. On the other hand, we find an even larger difference between the radial distributions of LGG
SBGGs and of the galaxies in the control sample ($p < 0.01$).

\begin{table*}
 \caption{Correlation analysis of residuals of BGG properties, fit to their velocity dispersions, with magnitude gap.
 The results for the whole sample and for a subsample of BGGs with $10\% < f_{\rm L} < 20\%$. {\bf (1)} Galaxy property; {\bf (2)} 
 Kendall correlation coefficients; {\bf (3)} $p$-values; and {\bf (4)} fraction $f$ of the $1000$ realizations of 
 the fit (see Section \ref{Sec_fit_relations}) that leads to $p < 0.05$. Columns {\bf (5, 6, 7)} show the same for the Spearman correlation test. }
\begin{tabular}{llrrrrrrr}
\hline
 &  &  \multicolumn{3}{c}{Kendall} & & \multicolumn{3}{c}{Spearman} \\
 \cline{3-5} \cline{7-9} 
 \multicolumn{2}{c}{BGG property} & \multicolumn{1}{c}{$\tau$} & \multicolumn{1}{c}{$p$-value} &  \multicolumn{1}{c}{$f$} & & 
            \multicolumn{1}{c}{$\rho$} & \multicolumn{1}{c}{$p$-value} &  \multicolumn{1}{c}{$f$} \\
            
 \multicolumn{2}{c}{(1)}  & \multicolumn{1}{c}{(2)} & \multicolumn{1}{c}{(3)} & \multicolumn{1}{c}{(4)}  & &   
                              \multicolumn{1}{c}{(5)} & \multicolumn{1}{c}{(6)} & \multicolumn{1}{c}{(7)}  \\                   
\hline
& & \multicolumn{7}{c}{Full sample ($550$ BGGs)} \\
&   $(g-i)$                              & $-0.01 \pm 0.01$  &  $0.60$  &  $ 0.0\%$  &  &  $-0.02 \pm 0.01$  &  $0.57$  &  $ 0.0\%$  \\   
& {\tt eClass}                           & $ 0.02 \pm 0.01$  &  $0.42$  &  $ 0.0\%$  &  &  $ 0.04 \pm 0.01$  &  $0.43$  &  $ 0.0\%$  \\   
& $\log (\rm{sSFR/yr})$                  & $ 0.00 \pm 0.01$  &  $0.88$  &  $ 0.0\%$  &  &  $ 0.01 \pm 0.01$  &  $0.86$  &  $ 0.0\%$  \\ 
\vspace{0.2cm}                                                                                    
& $[\alpha/{\rm Fe}]$                    & $ 0.02 \pm 0.01$  &  $0.44$  &  $ 0.0\%$  &  &  $ 0.03 \pm 0.01$  &  $0.44$  &  $ 0.0\%$  \\ 
\multicolumn{1}{l}{\multirow{4}{*}{BC03}}                                                         
&\multicolumn{1}{|l}{\ageFifty}          & $-0.02 \pm 0.01$  &  $0.39$  &  $ 0.0\%$  &  &  $-0.04 \pm 0.01$  &  $0.39$  &  $ 0.0\%$  \\   
&\multicolumn{1}{|l}{\ageNinety}         & $-0.06 \pm 0.01$  &  $0.05$  &  $13.3\%$  &  &  $-0.08 \pm 0.01$  &  $0.05$  &  $15.6\%$  \\   
&\multicolumn{1}{|l}{\ageFiftyNinety}    & $ 0.05 \pm 0.01$  &  $0.09$  &  $14.5\%$  &  &  $ 0.07 \pm 0.02$  &  $0.09$  &  $14.7\%$  \\   
\vspace{0.2cm}                                                                                     
&\multicolumn{1}{|l}{$[$M/H$]$}          & $ 0.05 \pm 0.01$  &  $0.11$  &  $ 0.7\%$  &  &  $ 0.07 \pm 0.01$  &  $0.10$  &  $ 0.8\%$  \\   
\multicolumn{1}{l}{\multirow{4}{*}{V15}}                                                           
&\multicolumn{1}{|l}{\ageFifty}          & $ 0.01 \pm 0.01$  &  $0.74$  &  $ 0.0\%$  &  &  $ 0.01 \pm 0.01$  &  $0.74$  &  $ 0.0\%$  \\   
&\multicolumn{1}{|l}{\ageNinety}         & $-0.01 \pm 0.01$  &  $0.78$  &  $ 0.0\%$  &  &  $-0.01 \pm 0.01$  &  $0.82$  &  $ 0.0\%$  \\   
&\multicolumn{1}{|l}{\ageFiftyNinety}    & $ 0.05 \pm 0.01$  &  $0.08$  &  $ 8.5\%$  &  &  $ 0.07 \pm 0.01$  &  $0.08$  &  $ 9.2\%$  \\   
\vspace{0.2cm}                                                                                    
&\multicolumn{1}{|l}{$[$M/H$]$}          & $ 0.02 \pm 0.01$  &  $0.44$  &  $ 0.4\%$  &  &  $ 0.03 \pm 0.01$  &  $0.45$  &  $ 0.4\%$  \\     
                                                  
& & \multicolumn{7}{c}{BGGs with $10\% < f_{\rm L} < 20\%$ (368 BGGs)} \\
&   $(g-i)$                              & $-0.01 \pm 0.01$  &  $0.70$  &  $0.0\%$  &  &  $-0.02 \pm 0.01$  &  $0.66$  &  $0.0\%$  \\   
& {\tt eClass}                           & $ 0.03 \pm 0.01$  &  $0.34$  &  $0.0\%$  &  &  $ 0.05 \pm 0.01$  &  $0.35$  &  $0.0\%$  \\   
& $\log (\rm{sSFR/yr})$                  & $ 0.02 \pm 0.01$  &  $0.61$  &  $0.0\%$  &  &  $ 0.03 \pm 0.01$  &  $0.60$  &  $0.0\%$  \\   
\vspace{0.2cm}                             
& $[\alpha/{\rm Fe}]$                    & $ 0.02 \pm 0.01$  &  $0.54$  &  $0.0\%$  &  &  $ 0.03 \pm 0.01$  &  $0.55$  &  $0.0\%$  \\   
\multicolumn{1}{l}{\multirow{4}{*}{BC03}}
&\multicolumn{1}{|l}{\ageFifty}          & $ 0.01 \pm 0.01$  &  $0.88$  &  $0.0\%$  &  &  $ 0.01 \pm 0.02$  &  $0.90$  &  $0.0\%$  \\   
&\multicolumn{1}{|l}{\ageNinety}         & $-0.03 \pm 0.01$  &  $0.39$  &  $0.0\%$  &  &  $-0.05 \pm 0.02$  &  $0.38$  &  $0.0\%$  \\   
&\multicolumn{1}{|l}{\ageFiftyNinety}    & $ 0.02 \pm 0.01$  &  $0.63$  &  $0.0\%$  &  &  $ 0.02 \pm 0.02$  &  $0.64$  &  $0.0\%$  \\   
\vspace{0.2cm}                           
&\multicolumn{1}{|l}{$[$M/H$]$}          & $ 0.03 \pm 0.01$  &  $0.46$  &  $0.0\%$  &  &  $ 0.04 \pm 0.01$  &  $0.46$  &  $0.0\%$  \\   
\multicolumn{1}{l}{\multirow{4}{*}{V15}}   
&\multicolumn{1}{|l}{\ageFifty}          & $ 0.00 \pm 0.01$  &  $0.92$  &  $0.0\%$  &  &  $ 0.01 \pm 0.01$  &  $0.86$  &  $0.0\%$  \\   
&\multicolumn{1}{|l}{\ageNinety}         & $-0.02 \pm 0.01$  &  $0.60$  &  $0.0\%$  &  &  $-0.03 \pm 0.01$  &  $0.63$  &  $0.0\%$  \\   
&\multicolumn{1}{|l}{\ageFiftyNinety}    & $ 0.05 \pm 0.01$  &  $0.13$  &  $7.2\%$  &  &  $ 0.08 \pm 0.01$  &  $0.12$  &  $8.3\%$  \\   
\vspace{0.2cm}                           
&\multicolumn{1}{|l}{$[$M/H$]$}          & $-0.02 \pm 0.01$  &  $0.58$  &  $0.0\%$  &  &  $-0.03 \pm 0.02$  &  $0.56$  &  $0.0\%$  \\                       
  
\hline   
\end{tabular}
\label{Tab_tests_bgg}
\end{table*}

\begin{table*}
 \caption{Same as Table~\ref{Tab_tests_bgg}, but for the SBGGs.}
\begin{tabular}{llrrrrrrr}
\hline
 &  &  \multicolumn{3}{c}{Kendall} & & \multicolumn{3}{c}{Spearman} \\
 \cline{3-5} \cline{7-9} 
 \multicolumn{2}{c}{SBGG property} & \multicolumn{1}{c}{$\tau$} & \multicolumn{1}{c}{$p$-value} &  \multicolumn{1}{c}{$f$} & & 
            \multicolumn{1}{c}{$\rho$} & \multicolumn{1}{c}{$p$-value} &  \multicolumn{1}{c}{$f$} \\
            
 \multicolumn{2}{c}{(1)}  & \multicolumn{1}{c}{(2)} & \multicolumn{1}{c}{(3)} & \multicolumn{1}{c}{(4)}  & &   
                            \multicolumn{1}{c}{(5)} & \multicolumn{1}{c}{(6)} & \multicolumn{1}{c}{(7)}  \\                   
\hline
& & \multicolumn{7}{c}{Full sample (138 elliptical SBGGs)} \\
&   $(g-i)$                               &  $ 0.09 \pm 0.01$  &  $0.11$  &  $ 2.4\%$  &  &  $ 0.14 \pm 0.01$  &  $0.11$  &  $ 1.6\%$  \\   
& {\tt eClass}                            &  $-0.01 \pm 0.01$  &  $0.93$  &  $ 0.0\%$  &  &  $-0.01 \pm 0.02$  &  $0.92$  &  $ 0.0\%$  \\   
& $\log (\rm{sSFR/yr})$                   &  $\bm{0.13 \pm 0.01}$  &  $\bm{0.02}$  &  $\bm{93.8\%}$  &  &  $ \bm{0.19 \pm 0.02}$  &  $\bm{0.02}$  &  $\bm{91.3\%}$  \\
\vspace{0.2cm}                                                                                      
& $[\alpha/{\rm Fe}]$                     &  $ 0.02 \pm 0.01$  &  $0.77$  &  $ 0.0\%$  &  &  $ 0.03 \pm 0.02$  &  $0.75$  &  $ 0.0\%$  \\ 
\multicolumn{1}{l}{\multirow{4}{*}{BC03}}                                                           
&\multicolumn{1}{|l}{\ageFifty}           &  $-0.03 \pm 0.01$  &  $0.64$  &  $ 0.0\%$  &  &  $-0.03 \pm 0.02$  &  $0.69$  &  $ 0.0\%$  \\   
&\multicolumn{1}{|l}{\ageNinety}          &  $ 0.00 \pm 0.02$  &  $0.95$  &  $ 0.0\%$  &  &  $ 0.01 \pm 0.04$  &  $0.93$  &  $ 0.0\%$  \\   
&\multicolumn{1}{|l}{\ageFiftyNinety}     &  $-0.03 \pm 0.03$  &  $0.66$  &  $ 0.0\%$  &  &  $-0.03 \pm 0.04$  &  $0.70$  &  $ 0.0\%$  \\   
\vspace{0.2cm}                                                                                       
&\multicolumn{1}{|l}{$[$M/H$]$}           &  $ 0.06 \pm 0.01$  &  $0.27$  &  $ 0.0\%$  &  &  $ 0.09 \pm 0.02$  &  $0.28$  &  $ 0.0\%$  \\              
\multicolumn{1}{l}{\multirow{4}{*}{V15}}                                                             
&\multicolumn{1}{|l}{\ageFifty}           &  $ 0.04 \pm 0.01$  &  $0.47$  &  $ 0.0\%$  &  &  $ 0.07 \pm 0.02$  &  $0.41$  &  $ 0.0\%$  \\   
&\multicolumn{1}{|l}{\ageNinety}          &  $-0.03 \pm 0.01$  &  $0.62$  &  $ 0.0\%$  &  &  $-0.05 \pm 0.02$  &  $0.57$  &  $ 0.0\%$  \\   
&\multicolumn{1}{|l}{\ageFiftyNinety}     &  $ 0.08 \pm 0.01$  &  $0.17$  &  $ 0.0\%$  &  &  $ 0.12 \pm 0.02$  &  $0.15$  &  $ 0.0\%$  \\   
\vspace{0.2cm}                                                                                      
&\multicolumn{1}{|l}{$[$M/H$]$}           &  $ 0.07 \pm 0.01$  &  $0.22$  &  $ 0.1\%$  &  &  $ 0.10 \pm 0.02$  &  $0.24$  &  $ 0.0\%$  \\   
                                                  
& & \multicolumn{7}{c}{SBGGs with $20\% < f_{\rm L} < 30\%$  (63 elliptical SBGGs)} \\
&   $(g-i)$                               &  $ 0.14 \pm 0.03$  &  $0.11$  &  $18.5$  &  &  $ 0.18 \pm 0.05$  &  $0.16$  &  $12.4$ \\   
& {\tt eClass}                            &  $-0.13 \pm 0.02$  &  $0.14$  &  $ 4.3$  &  &  $-0.17 \pm 0.04$  &  $0.18$  &  $ 2.6$ \\   
& $\log (\rm{sSFR/yr})$                   &  $ 0.08 \pm 0.02$  &  $0.38$  &  $ 0.0$  &  &  $ 0.10 \pm 0.04$  &  $0.43$  &  $ 0.0$ \\   
\vspace{0.2cm}                               
& $[\alpha/{\rm Fe}]$                     &  $ 0.09 \pm 0.02$  &  $0.32$  &  $ 0.2$  &  &  $ 0.12 \pm 0.03$  &  $0.34$  &  $ 0.1$ \\   
\multicolumn{1}{l}{\multirow{4}{*}{BC03}}    
&\multicolumn{1}{|l}{\ageFifty}           &  $ 0.03 \pm 0.02$  &  $0.76$  &  $ 0.0$  &  &  $ 0.04 \pm 0.04$  &  $0.74$  &  $ 0.0$ \\   
&\multicolumn{1}{|l}{\ageNinety}          &  $ 0.10 \pm 0.03$  &  $0.24$  &  $ 1.0$  &  &  $ 0.16 \pm 0.04$  &  $0.22$  &  $ 1.4$ \\   
&\multicolumn{1}{|l}{\ageFiftyNinety}     &  $-0.11 \pm 0.03$  &  $0.21$  &  $ 1.4$  &  &  $-0.18 \pm 0.05$  &  $0.17$  &  $ 2.9$ \\   
\vspace{0.2cm}                               
&\multicolumn{1}{|l}{$[$M/H$]$}           &  $ 0.04 \pm 0.03$  &  $0.67$  &  $ 0.0$  &  &  $ 0.07 \pm 0.05$  &  $0.61$  &  $ 0.1$ \\   
\multicolumn{1}{l}{\multirow{4}{*}{V15}}     
&\multicolumn{1}{|l}{\ageFifty}           &  $ 0.13 \pm 0.02$  &  $0.14$  &  $ 2.6$  &  &  $ 0.21 \pm 0.03$  &  $0.09$  &  $ 7.3$ \\   
&\multicolumn{1}{|l}{\ageNinety}          &  $ 0.00 \pm 0.03$  &  $0.98$  &  $ 0.0$  &  &  $ 0.02 \pm 0.04$  &  $0.89$  &  $ 0.0$ \\   
&\multicolumn{1}{|l}{\ageFiftyNinety}     &  $ 0.14 \pm 0.03$  &  $0.11$  &  $ 0.6$  &  &  $ 0.20 \pm 0.04$  &  $0.13$  &  $ 0.2$ \\   
\vspace{0.2cm}                               
&\multicolumn{1}{|l}{$[$M/H$]$}           &  $ 0.05 \pm 0.02$  &  $0.59$  &  $ 0.0$  &  &  $ 0.07 \pm 0.04$  &  $0.61$  &  $ 0.0$ \\                   
  
\hline   
\end{tabular}
\label{Tab_tests_sbgg}
\end{table*}

%

\begin{table}
 \caption{Comparison between SBGGs in LGGs and the control sample. We applied the Propensity Score Matching analysis using
 {\bf (1)} \MstarNoUnit; {\bf (2)} \MstarNoUnit\ and \Mabs; {\bf (3)} \MstarNoUnit\ and morphology. We show the differences between the
 median values ($\Delta${\bf median}) and the quantile-quantile distances between the distributions ({\bf Q-Q distances}) of 
 \MstarNoUnit, \Mabs, \MvirNoUnit, and $R/r_{\rm vir}$ of the LGG SBGGs and control samples. The comparison between the properties of galaxies in the two samples ({\bf KS $\bm p$-values})  
 and the fraction of different {\bf morphological types} in the control sample are shown.}
\begin{tabular}{rrrrr}
\hline
                         &                  &    \MstarNoUnit   & \MstarNoUnit      & \MstarNoUnit          \\
                         &                  &                   & $\&$  \Mabs       & $\&$ morph.           \\
                         &                  &                   &                   &                       \\
                         &                  &       (1)         &        (2)        &       (3)             \\                 
\cline{3-5} \\ 
\multicolumn{2}{l}{$\Delta$ {\bf median}}   &  \multicolumn{3}{c}{  }                                       \\                                               
\multicolumn{2}{r}{\Mstar      }            &  $\emph{0.0005}$  &  $-\emph{0.006}$  &  $\emph{0.020}$       \\      
\multicolumn{2}{r}{\Mabs\ (mag)}            &       $ -0.045$   &  $  \emph{0.01}$  &  $-0.053$             \\     
\multicolumn{2}{r}{\Mvir       }            &       $ -0.016$   &        $-0.019$   &  $-0.015$             \\   
\multicolumn{2}{r}{$f_{\rm L}~(\%)$}        &       $  -2.05$   &        $ -2.77$   &  $ -1.57$             \\     
\multicolumn{2}{r}{$R/r_{\rm vir}$}         &       $  0.053$   &        $ 0.062$   &  $ 0.073$             \\     

\multicolumn{2}{l}{{\bf Q-Q distances}}     &  \multicolumn{3}{c}{  }                                       \\     
\multicolumn{2}{r}{\Mstar      }            &$-\emph{0.0003}$   &  $-\emph{0.014}$  &  $-\emph{0.0002}$     \\     
\multicolumn{2}{r}{\Mabs\ (mag)}            &      $ -0.0022$   &  $ \emph{0.018}$  &  $-0.0052$            \\     
\multicolumn{2}{r}{\Mvir       }            &      $  0.0085$   &        $-0.002$   &  $  0.008$            \\ 
\multicolumn{2}{r}{$f_{\rm L}~(\%)$}        &      $   -1.48$   &        $ -2.53$   &  $  -1.33$             \\     
\multicolumn{2}{r}{$R/r_{\rm vir}$}         &      $  0.0425$   &        $ 0.043$   &  $  0.052$            \\     
                                                                                             
\multicolumn{2}{l}{{\bf KS $\bm p$-values}} &  \multicolumn{3}{c}{  }                                       \\     
\multicolumn{2}{r}{  $(g-r)$   }            &     $ 0.08$       &     $ 0.13$       & $ 0.07$               \\     
\multicolumn{2}{r}{  $(g-i)$   }            &     $ 0.29$       &     $ 0.41$       & $ 0.41$               \\     
\multicolumn{2}{r}{{\tt eClass}}            &     $ 0.63$       &     $ 0.40$       & $ 0.72$               \\     
\multicolumn{2}{r}{$\log (\rm{sSFR/yr})$}   &     $ 0.29$       &     $ 0.50$       & $ 0.23$               \\     
                  \\
\multicolumn{1}{l}{\multirow{4}{*}{BC03}}
&\multicolumn{1}{|r}{\ageFifty}             &     $ 0.23$       &     $ 0.07$       & $ 0.13$               \\     
&\multicolumn{1}{|r}{\ageNinety}            &     $ 0.65$       &     $ 0.50$       & $ 0.75$               \\                                                                             
&\multicolumn{1}{|r}{\ageFiftyNinety}       &     $ 0.60$       &     $ 0.29$       & $ 0.75$               \\
&\multicolumn{1}{|r}{$[$M/H$]$}             &     $ 0.80$       &     $ 0.50$       & $ 0.92$               \\     
                    \\                                            
\multicolumn{1}{l}{\multirow{4}{*}{V15}}                                                                                                 
&\multicolumn{1}{|r}{\ageFifty}             &     $ 0.60$       &     $ 0.70$       & $ 0.60$               \\     
&\multicolumn{1}{|r}{\ageNinety}            &     $ 0.15$       &     $ 0.11$       & $ 0.15$               \\                                                                              
&\multicolumn{1}{|r}{\ageFiftyNinety}       &     $ 0.80$       &     $ 0.60$       & $ 0.89$               \\
&\multicolumn{1}{|r}{$[$M/H$]$}             &     $ 0.89$       &     $ 0.85$       & $ 0.45$               \\     
                           \\
&\multicolumn{1}{r}{$R/r_{\rm vir}$}        &  $\bm{0.020}$     &   $\bm{0.007}$    & $\bm{0.001}$          \\     
\\
\multicolumn{2}{l}{\bf Morphological types} &  \multicolumn{3}{c}{  }                                       \\                                                                     
\multicolumn{2}{r}{$f_{\rm elliptical}$}    &    $  8.5\%$      &    $ 10.9\%$      &    $  9.4\%$          \\  
\multicolumn{2}{r}{$f_{\rm spiral}$    }    &    $ 25.5\%$      &    $ 26.4\%$      &    $ 20.8\%$          \\
\multicolumn{2}{r}{$f_{\rm uncertain}$ }    &    $ 65.1\%$      &    $ 61.8\%$      &    $ 69.8\%$          \\

\hline    
\end{tabular}
\label{Tab_PSM}
\end{table}

\section{Discussion}
\label{Sec_discussion}

\subsection{Stellar populations of the brightest group galaxies}
\label{Sec_mergers_vs_SFH}

As shown in Figs.~\ref{Fig_dColor_dMag_sigma} to \ref{Fig_dAgeMet_dMag_miles_sigma} and in Table~\ref{Tab_tests_bgg}, we find no significant
trends of BGG properties with gap, suggesting that all 
BGGs share the same stellar population properties, regardless of the magnitude gap.

The absence of significant variations of the stellar population properties
with magnitude gap is in agreement with previous results, 
e.g. \cite{LaBarbera.etal:2009}, \cite{Harrison.etal:2012}, and \cite{Eigenthaler.Zeilinger:2013}. These authors
compared the stellar populations (age, metallicities, [$\alpha$/Fe], colours,
and the radial variation of these properties) of BGGs in LGGs and normal
groups, and they found no distinction between the LGGs and the control
samples.

This lack of variation of SFH with magnitude gap suggests that
all BGGs are formed in a very similar way, regardless of the magnitude gap.
However, at a fixed halo mass, BGGs in LGGs are more massive than their
counterparts in SGGs (Fig.~\ref{Fig_fossils_E}a), which could be due to a
higher star formation efficiency in LGG BGGs compared to those in SGGs. 
But, in this case, LGG BGGs should have higher metallicities than BGGs in SGGs, which we do not observe in our analysis once we correct for 
galaxy velocity dispersion.
Alternatively, they may have formed the 
bulk of their stellar masses in an early wet merger event, which would have little effect on the traceable SFHs of these systems.
Finally, LGG BGGs could have undergone more
mergers than BGGs in SGGs. In this scenario, the fact that we do not 
observe any trend with gap of ages and metallicities implies that all BGGs would have 
been formed by mergers, and these mergers were dry.

The analysis by \citet{DiazGimenez.etal:2008} of galaxies in semi-analytical models run on the Millenium simulations show that BGGs in haloes more massive than $5 \times 10^{13}$~M$_{\odot}$ are mainly formed by gas-poor mergers, regardless of the magnitude gap of their host haloes. 
The median redshift to form half of the stellar mass is $z \sim
3.5$ for central galaxies in LGGs and $z \sim 3.7$ in SGGs, which
corresponds to median ages of $\approx 11.8$ and $12$~Gyr.
However, the difference of $\sim 0.2\,\rm Gyr$ in median age is not  detectable
with the time resolution of our SFH analysis. 

\citeauthor{DiazGimenez.etal:2008} furthermore show that, although LGGs assembled most of their virial mass at higher redshifts in comparison with SGGs, BGGs in LGGs merged later compared to their non-LGG counterparts: the last major merger in LGGs and SGGs ocurred $\sim 4.3$~Gyr and $\sim 4.7$~Gyr ago (median), respectively. 
The stellar population synthesis (SPS) method can trace the SFHs, but not the merger
history, i.e., it is possible to determine when the stars were formed but not
when they were accreted to the BGG.  However, a late wet merger followed by a burst of
star formation could be identified in the SFH of a galaxy. 
But the typical time difference between the last major mergers in LGGs vs SGGs of  $0.4$~Gyr at a lookback time of over $4$~Gyr ago is
still challenging to detect in an SPS analysis.

Our results also show that the higher
$M_{\rm halo}/L_r$ ratios in LGGs compared to that in LGGs found in several studies 
\citep{Jones.etal:2003, Yoshioka.etal:2004, Khosroshahi.etal:2007, Proctor.etal:2011} 
are unlikely to be due to differences in the stellar population properties.
Since we find no variation of stellar population properties with gap 
(i.e., $M_{\star}/L_r$ is independent of \deltam) and since the halo mass-to-luminosity ratios can be written as $M_{\rm halo}/L_r = (M_{\rm halo}/M_{\star}) (M_{\star}/L_r)$, 
then any variation of $M_{\rm halo}/L_r$ with gap must be a consequence of differences in the $M_{\star}/M_{\rm halo}$ ratios.
Therefore, the high halo mass-luminosity ratios of LGGs can be interpreted as low $M_{\star}/M_{\rm halo}$ ratios, as previously suggested.
Alternatively, the halo mass-luminosity ratios of LGGs could be, in fact, no different from that of SGGs, 
as suggested in other studies \citep[e.g.,][]{Voevodkin.etal:2010, Harrison.etal:2012, Girardi.etal:2014}. 
In a forthcoming paper \citep{Trevisan.etal:2016b}, we will address the $M_{\star}/L_r$ and the $M_{\star}/M_{\rm halo}$ ratios versus gap 
relations in more detail.

Finally, we repeated the analysis presented in Section~\ref{Sec_m12_StarPop}
using stellar masses instead of velocity dispersions, and we find no statistically significant trend of the stellar mass-corrected properties with \deltam\ for any of the BGG and the SBGG properties.
In addition, we fitted the properties versus \veldispNoUnit\ relations using a second-order polynomial to investigate if a different fitting method affects our results.
We also changed the number of bins by choosing bins of $50$ galaxies instead of $100$.
In both cases, the resulting Kendall and Spearman correlation coefficients are very similar to those that we obtain when we fit the relations with LOESS
using bins with $100$ galaxies (see Section~\ref{Sec_fit_relations}). 
Again, we find no significant trends of the residuals of stellar population
diagnostics with magnitude gap, except, again for sSFR in SBGGs, but this
trend is no longer significant (again) when we limit the range of the
fraction of the luminosity within the fibre.

%
\begin{figure*}
\centering
\includegraphics[width=\hsize]{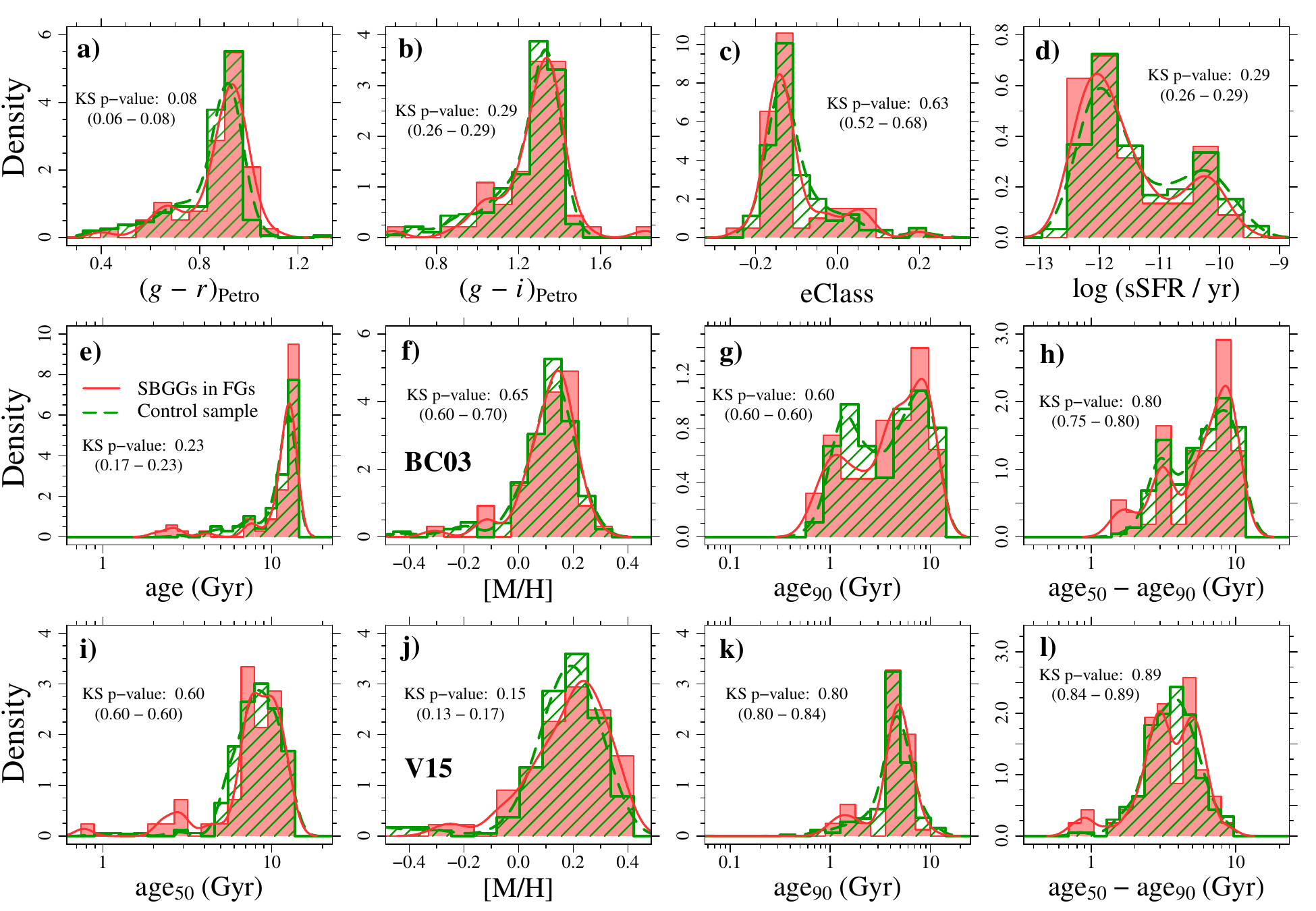}
\caption {Comparison between the properties of second-ranked galaxies in
  LGGs (\emph{red}) and the control sample in groups with \deltam~$< 2.0$~mag
  (\emph{green}), 
  described in Section \ref{Sec_PSM}. 
  {\it Upper panels:} $g-r$ ({\bf a}) and $g-i$ ({\bf b})
  colours, {\tt eClass} parameters ({\bf c}), and sSFR ({\bf d}). Second-ranked galaxies in LGGs
  groups and the control sample are respresented by red and green
  lines/histograms, respectively.
  {\it Middle and lower panels:} median ages ({\bf e, i}),
  metallicities ({\bf f, j}), lookback times when $90$\% of the stellar mass was formed
  ({\bf g, k}), and elapsed time from the formation of $50\%$ to $90\%$ of the
  total stellar mass formed in the galaxy ({\bf h, l}). 
The quantities derived from spectral modeling based on
the BC03 SSP model are presented in panels e to h, and panels i to l show the
results obtained with the V15 model.
In each panel we indicate the $p$-value of the KS test (median, $1$ and $99$-percentiles).
} 
\label{Fig_rank2_control}
\end{figure*}

\subsection{Projected separation between BGGs and SBGGs versus gap}
\label{Sec_discussion_sbgg}

The comparison between SBGGs and normal
galaxies of all morphological types (Section~\ref{Sec_PSM}) shows that the
stellar populations of SBGGs in LGGs are statistically compatible with similar galaxies
in normal groups (Figs.~\ref{Fig_rank2_control}). 

\begin{figure}
\centering
\includegraphics[width=0.9\hsize]{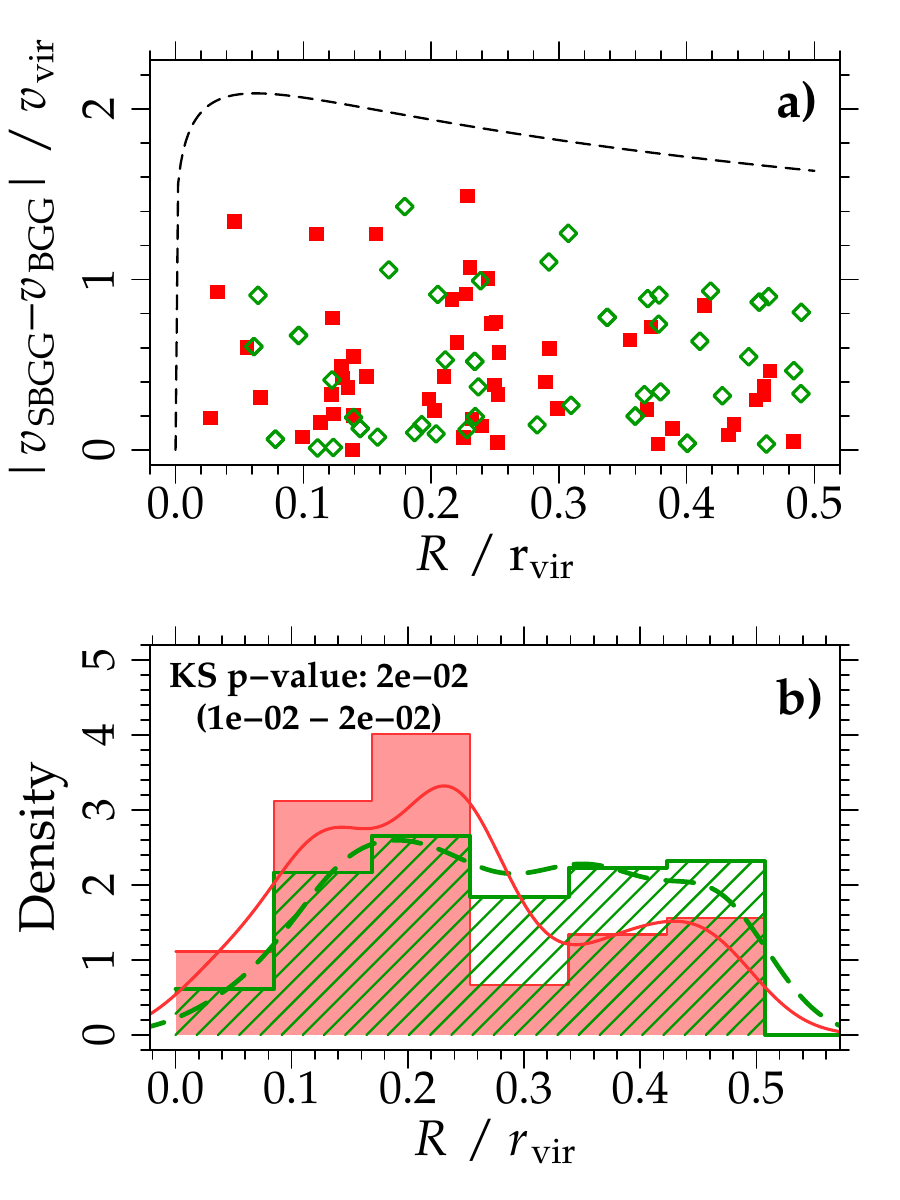}
\caption {Diagram of $|v_{\rm los}| / v_{200}$ vs. $R / {\rm r}_{\rm vir}$ (\emph{upper panel}) and 
normalized projected distance to the BGG {\emph{lower panel}}.
The second-ranked galaxies in LGGs and the control sample in groups with \deltam~$< 2.0$~mag
  (\emph{green}), described in Section \ref{Sec_PSM} are shown in \emph{red} and \emph{green}, respectively.
  In panel {\bf b} we indicate the $p$-value of the KS test (median, 1st and
  99th percentiles)}
\label{Fig_Rrvir_sbgg}
\end{figure}

However, we find that SBGGs in LGGs lie significantly closer to the BGGs (in
projection, Fig.~\ref{Fig_Rrvir_sbgg}). Dynamical friction should cause orbits to decay faster for galaxies whose
subhalo masses are greater in terms of the halo (group) mass. But since our
control sample was designed to have the same set of stellar masses as our LGG
SBGGs, and since we found no trend of lower halo mass for the SBGG LGGs
(Fig.~\ref{Fig_PSM}c), one does not expect to have different orbital decay
times.
Therefore, the lower normalized radii of SBGGs in LGGs must indicate that
these galaxies entered their groups at earlier times than the similar stellar
mass galaxies of the control sample.

To estimate the time of entry, one can assume that the mean density scales as $r^{-2}$ near the group scale radius ($\sim r_{\rm vir} / 4$),
and also as $t^{-2}$ (which is correct for $\Omega_{\rm M} = 1$ and $\Omega_{\Lambda} = 0$). Therefore, the time of entry 
scales roughly as the radius. According  to Table~\ref{Tab_PSM} and Fig.~\ref{Fig_Rrvir_sbgg}b, the SBGGs in LGGs lie $\Delta r \sim 0.05~r_{\rm vir}$ closer to the BGG than similar galaxies in normal
groups, which means that they entered the group $\Delta t \sim \Delta r\ t/r$ earlier. Assuming the NFW profile at all times and that the physical density remains constant within the virial radius 
\citep[see][]{Mamon:1992}, then we are able to compute the time of entry of a galaxy into a group through the relation 
\begin{equation}
 \rho_{\rm mean}(r, t = t_0) = \rho_{\rm mean}(r_{\rm vir} = r, t) \nonumber
\end{equation}
which leads to solving 
\begin{equation}
 \Delta(z) H^2(z) =  \Delta(z= 0 ) H_0^2\ \frac{\ln(c\ x + 1) - \left[c\ x/(c\ x + 1) \right]}{\ln (c) - \left[c / (c + 1)\right] } 
 \label{Eq_tentry}
\end{equation}
where $x = r/r_{\rm vir}(z\!=\!0)$ and $c$ is the concentration. Solving eq.~(\ref{Eq_tentry}) by shifting $x$ by $-0.05~r_{\rm vir}$ and comparing the results, we obtain $\Delta t = 600$~Myr.

One may wonder whether the virial radii of LGGs could have been overestimated, leading to lower
$R/r_{\rm vir}$ values for SBGGs within these groups. However, the group masses determined by \citeauthor{Yang.etal:2007} catalogue (hence the radii derived from them) are not expected to be
biased by the magnitude gap, since the abundance matching between observed groups and the theoretical halo mass function is performed using the total 
group luminosity. On the other hand, if LGGs really have abnormally high  $M_{\rm halo}/L_r$, as suggested by many authors 
(\citealp{Jones.etal:2003, Yoshioka.etal:2004, Khosroshahi.etal:2007, Proctor.etal:2011}; but see \citealp{Voevodkin.etal:2010, Harrison.etal:2012, Girardi.etal:2014} for an alternative view),
\citeauthor{Yang.etal:2007} could have underestimated the halo masses of LGGs, since they derive halo masses from group optical luminosities with abundance matching.
Therefore, the group virial radii may have been underestimated for the LGGs, and hence the normalized projected distances between the SBGGs and BGGs in LGGs may be even smaller, meaning that SBGGs in LGGs may have entered the group more than $600$~Myr earlier than similar galaxies in regular groups.

If the earlier entry scenario is correct, and given the known segregation of fraction of quenched (or inversely of
star-forming) galaxies (e.g., \citealp{vonderLinden.etal:2010, Mahajan.etal:2011}), 
one would expect that the radial segregation of SBGG in LGGs relative to galaxies of the
same stellar mass in regular groups would lead to the former having older
stellar populations. 
Nevertheless, we do not observe any difference between the ages of LGG SBGGs and those of similar galaxies in regular groups. 
In addition, the fractions of star-forming galaxies ($\log [{\rm sSFR/yr}] > -11$) among the LGG SBGGs and galaxies in the control sample
are very similar ($25\%$ and $34\%$, respectively), with Barnard's test indicating low statistical significance ($p > 0.1$).

%

\subsection{Fossil versus large-gap groups}

In the formal definition \citep{Jones.etal:2003}, a group is classified as
fossil if its bolometric X-ray luminosity is greater than $10^{42} h^{-2}_{50}$~erg~s$^{-1}$ (in addition to the \deltam~$> 2$~mag within $0.5\,r_{\rm vir}$). Therefore, since our sample definition does not include the X-ray criteria, some of our LGGs might not be FGs\footnote{However, we note that according to \cite{Dariush.etal:2007}, groups with \Mvir~$\gtrsim 13.2$ are typically X-ray FGs.}. 

Fossil groups are believed to be systems that have assembled their masses at
relatively early times. However, according to \citet{Raouf.etal:2014}, who
analyzed a semi-analytical model of galaxy formation, only a fraction of
large-gap groups are in fact ``old'' systems, i.e., assembled at least half
of their total mass at $z > 1$. \citeauthor{Raouf.etal:2014} 
predict that $\sim\!85\%$ of groups with $2 < $~\deltam~$ < 2.5$ and $r$-band luminosities between $1.9 \times 10^{10}$ and $2.0 \times 10^{11}~h^{-2}~{\rm L}_{\odot}$ are ``old'' groups. The fraction of old systems decreases to $\sim 50\%$ among high luminosity groups ($2.0 \times 10^{11} < L_r < 5.7 \times 10^{11}~h^{-2}~{\rm L}_{\odot}$). 

Given that $54$ ($91\%$) of our large-gap groups have luminosities $< 2 \times 10^{11}~h^{-2}~{\rm L}_{\odot}$, and all of them have are less luminous than $5 \times 10^{11}~h^{-2}~{\rm L}_{\odot}$, we estimate the fraction of old groups (i.e., true FGs) in our sample to be $\sim 80\%$ according to the predictions of \citeauthor{Raouf.etal:2014}.
However, since the assembly history of haloes cannot be directly observed, it is very difficult to compare predictions from simulations and observations. 
So, although a connection between LGGs and FGs must exist, determining the exact relation between these two classes of systems is challenging. 
Hence, it is not clear whether our conclusions on the lack of differences in the SFHs of BGGs and SBGGs within LGGs and SGGs may be generalized to the $1^{\rm st}$ and the $2^{\rm nd}$-ranked galaxies in fossil vs. non-fossil groups.

\section{Summary and conclusions}
\label{Sec_summary}

In this study, we used a complete sample of $550$ SDSS groups to investigate how
the properties and star formation histories of the BGGs (restricted to elliptical morphologies) and SBGGs vary with the magnitude gap,  
after removing the dependences with velocity dispersion and stellar mass. 
We computed galaxy median ages,
the lookback times at which $90$\% of the total stellar mass was formed, mass-weighted metallicities, and \alphaFe.
We also examined galaxy colours, specific star formation rates, and the {\tt eClass} parameter. 

While the trends of stellar populations with velocity dispersion (or stellar mass) often show major differences according to the chosen single stellar population model, 
several conclusions stand out, all of which are independent of the adopted spectral model:

\begin{description}

\item[--] After correcting for trends with velocity dispersion, the BGGs follow the same distributions of $g-i$ colour, {\tt eClass}
  values, sSFRs, \alphaFe\ (Fig.~\ref{Fig_dColor_dMag_sigma}e--h), ages, metallicities, and SFHs derived with both the \citet{Vazdekis.etal:2015} and the \citet{Bruzual.Charlot:2003} models
  (Figs.~\ref{Fig_dAgeMet_dMag_bc03_sigma}e--h and \ref{Fig_dAgeMet_dMag_miles_sigma}e--h), regardless of the magnitude gap of
  their host group. We analysed a subsample of BGGs with similar fraction of their total luminosity within the aperture of the SDSS fibre (Table~\ref{Tab_tests_bgg}), and still no trend of BGG properties with \deltam\ is observed.  \\
 
\item[--] We found that elliptical SBGGs in groups with large gaps are very similar to those in small-gap groups, 
  and the analysis of an homogeneous sample of SBGGs shows that there are no significant trends of their properties with gap. \\ 

\item[--] Similarly, the stellar population properties of SBGGs of all morphologies in groups with \deltam~$\ge 2.0$~mag are very similar to the general population of galaxies with similar stellar masses (Fig.~\ref{Fig_rank2_control}). \\
 
\item[--] The projected separation between SBGGs and BGGs is smaller in groups with large gaps than 
 galaxies with similar stellar masses and magnitudes residing in normal groups (Fig.~\ref{Fig_Rrvir_sbgg}). 
 This appears to be due to the earlier entry of SBGGs within their now large-gap groups by $\sim 600$~Myr compared to similar galaxies in normal groups.
 
\end{description}


In a companion paper \citep*{Trevisan.etal:2016b}, we shed light on the merger
histories of brightest group galaxies, thus constraining both their mass
assembly histories and star formation histories. 

\section*{Acknowledgments}

The authors thank the anonymous referee for very detailed, thoughtful, and constructive comments that led to significant improvements in our manuscript.
MT acknowledges financial support from CNPq (process \#204870/2014-3). 
MT acknowledges T.~C.~Moura for kindly providing us with the spectral indices measurements used to obtain our \alphaFe\ estimates (Section~\ref{Sec_alpha}).
MT thanks the COIN collaboration (\url{https://asaip.psu.edu/organizations/iaa/iaa-working-group-of-cosmostatistics}) for 
providing the script to apply the PSM to our data. The preliminary version of the script was developed during the 2$^{\rm nd}$ COIN Residence Program (\url{http://iaacoin.wix.com/crp2015}); more details can be found in \citet{deSouza.etal:2016}.
We acknowledge the use of SDSS data (\url{http://www.sdss.org/collaboration/credits.html}) and {\tt TOPCAT} Table/VOTable Processing Software \citep[][\url{http://www.star.bris.ac.uk/~mbt/topcat/}]{topcat}. 

\appendix

\onecolumn

\section{Conversion from Yang $r_{180,m}$ to $r_{200,c}$}
\label{Ap_mean_to_critical}

The Yang et al. (2007) group catalogue provides group masses defined at the
radius where the mean density is $\Delta_{\rm Y} = 180$ times the mean
density of the Universe, $\overline\rho_{\rm U}(z)$,
at the group redshift. Thus, at $z=0$ and with $\Omega_{\rm m,0}=0.238$
adopted by Yang et al., the Yang group virial radius,
$r_{\rm Y} \equiv r_{\rm vir,Y}$
corresponds to a mean density of $180\,\Omega_{\rm m,0} \simeq 42.8$ times the
critical density of the Universe, $\rho_{\rm c,0} \equiv \rho_{\rm c}(0)$, where
\begin{equation}
\rho_{\rm c}(z) = {3\,H^2(z)\over 8\pi\,G} \ .
\label{rhocrit}
\end{equation}
The Yang group mean density can be written 
\begin{equation}
\overline \rho_{\rm Y} = 180\,\overline \rho_{\rm U}(z)  = 180\,\Omega_{\rm
  m}(z)\,\rho_{\rm c}(z)
= 180\, \Omega_{\rm m,0}\,(1+z)^3\,\rho_{\rm c,0} \ .
\label{rhomeanYang}
\end{equation}

We now define the groups at the radius $r_{200}$ where the mean density is 200
times the critical density
\begin{equation}
\label{rhomeannew}
\overline \rho_{200} = 200\,\rho_{\rm c}(z)
\ ,
\end{equation} 
so that the mass within the virial radius is
\begin{equation}
M_{200} = 100\,{H^2(z) \,r_{200}^3\over G}  \ .
\label{m200}
\end{equation}

{}From equations~(\ref{rhomeanYang}) and (\ref{rhomeannew}),  the ratio of mean densities is
\begin{equation}
{\overline \rho_{\rm Y} \over \overline \rho_{200}} = 0.9\,\Omega_{\rm
  m,0}\,{(1+z)^3\over E^2(z)}
\ ,
\label{rhomeanratio}
\end{equation}
where $E(z) = H(z)/H_0 = \sqrt{\Omega_{\rm m}^0 (1+z)^3 + 1-\Omega_{\rm m}}$
for a flat Universe. 

Assuming an NFW model (Navarro et al. 1996) for the mass distribution in the groups, 
the mean density profile can be expressed as
\begin{equation}
\overline \rho(r) = \overline\rho(a)\,\widetilde \rho\left ({r\over
  a}\right)
= \overline\rho(a)\,{\widetilde M(r/a)\over (r/a)^3} \ ,
\label{rhomeannfw}
\end{equation}
where
\begin{eqnarray} 
\widetilde \rho(x) &=& {\widetilde M(x) \over x^3} 
\label{rhomeantildenfw}\\
\widetilde M(x) &=& {M(a\,x)\over M(a)} = {\ln(x+1)-x/(x+1)\over \ln 2 - 1/2}
\label{mtildenfw}
\ ,
\end{eqnarray}
with $\widetilde \rho(1) = \widetilde M(1) = 1$.
Since the scale radius $a$ is fixed, 
equations~(\ref{rhomeanratio}) and (\ref{rhomeannfw}) lead to
\begin{equation}
{\overline \rho_{\rm Y} \over \overline \rho_{200}} 
= {\widetilde \rho (c_{\rm Y})
\over \widetilde \rho (c_{200}) } = 0.9\,\Omega_{\rm m,0}\,{(1+z)^{3}\over E^2(z)} \ ,
\end{equation}
where $c_{200} = r_{200}/a$ and
 $c_{\rm Y} = r_{\rm Y}/a$.
We can use the concentration-mass relation for $\Lambda$CDM halos (that of
Maccio et al. 2008), $c_{200} = c_{\Lambda \rm CDM}(M_{200})$, and
we  write the Yang concentration parameter as 
\begin{eqnarray}
c_{\rm Y} = c_{200}\,{r_{\rm Y}\over r_{200}} &=& 
 c_{\rm \Lambda CDM}(M_{200})\,\left[\left ({M_{\rm Y}\over M_{200}}\right)
\left({\overline \rho_{200} \over \overline \rho_{\rm Y}}\right)
\right]^{1/3} \nonumber \\
&=& {200\over180\,\Omega_{m,0}}\,
{c_{\Lambda \rm CDM}(M_{200})\,E^{2/3}(z)\over 1+z}\,\left ({M_{\rm Y}\over
  M_{200}}\right)^{1/3}\ .
\label{cYang}
\end{eqnarray}
Combining equations~(\ref{rhomeanratio}) and (\ref{cYang}), one can solve
\begin{equation}
\widetilde M 
\left(
{(0.9\,\Omega_{\rm m}^0)^{-1/3}\,c_{\rm \Lambda CDM}(M_{200})
\left [E^{2/3}(z)/(1+z)\right]
\left(M_{\rm Y}/ M_{200}\right)^{1/3}\over 
\widetilde M\left[c_{\rm \Lambda CDM}(M_{200}) \right]
}
\right)
\,M_{200} = M_{\rm Y}
\end{equation}
for $M_{200}$, where we used
using equations~(\ref{rhomeantildenfw}) and (\ref{mtildenfw}), where the $(\ln
2 -1/2)$ term in the latter scales out. 

 
The virial radius $r_{200}$ is then obtained by inverting
equation~(\ref{m200}) to give
\begin{equation}
r_{200} = \left ( {1\over 100}\,{G\,M_{200} \over H_0^2} \right)^{1/3}
\simeq 432\,\left ({M_{200}\over 10^{13}\,{\rm M}_\odot}\right)^{\textcolor{red}{1/3}}\, \rm kpc 
\end{equation}
for $H_0 = 73 \,\rm km \,s^{-1} \, Mpc^{-1}$, as adopted by Yang et al..

\twocolumn

\section{Propensity Score Matching}
\label{Ap_PSM}
%
%
\begin{figure}
\centering
\includegraphics[width=\hsize]{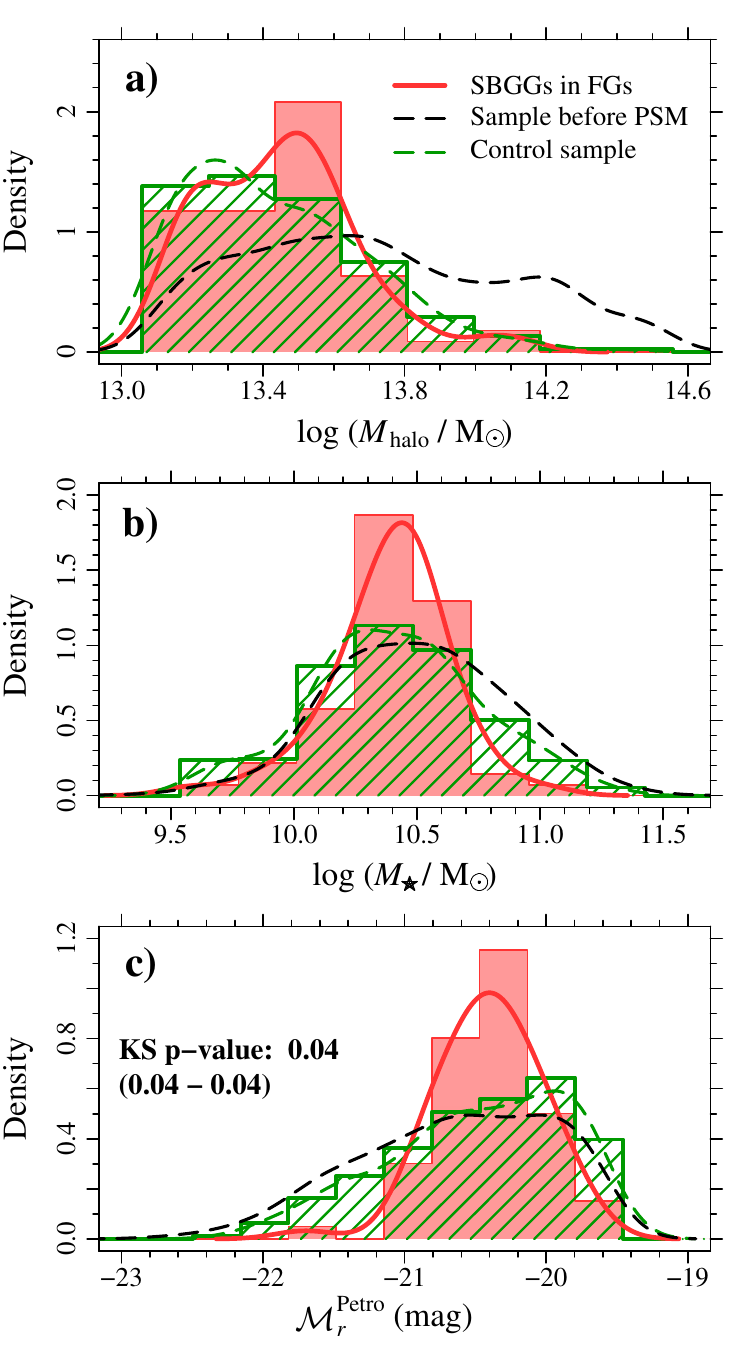}
\caption {Halo masses ({\it top}), stellar masses ({\it middle}) and
  absolute magnitudes ({\it bottom}) before and after the PSM (see  for
  details). The red histograms show the distributions
  of \Mabs, \MstarNoUnit, and \MvirNoUnit~of second-ranked galaxies in groups with
  \deltam~$> 2.0$~mag. The properties of the general population of galaxies in groups with \deltam~$<
  2$~mag (excluding the BGGs) are indicated by the black
  histograms. The control sample, drawn from the
  sample of galaxies in groups with \deltam~$< 2$~mag by applying the PSM
  technique (using both halo mass and stellar mass to perform the matching), is shown as the \emph{green histograms}.
The curves are obtained by smoothing the positions of the data points (not
the histograms) using a Gaussian kernel with standard deviation equal to one third
of the standard deviation of the data points. 
In panel {\bf c}, we indicate the $p$-value of the KS test (median, $1$ and $99$-percentiles).}
\label{Fig_PSM_ap}
\end{figure}


To apply the PSM technique, we used the R package {\sc MatchIt} \citep{MatchIt:2011}. The main goal is to select from the ``\emph{untreated sample}'' a control sample in which the distributions of observed properties are as similar as possible to those of the ``\emph{treated sample}''. 
First, the propensity scores $p_k$ (the probability that the unit $k$ will receive treatment) are estimated. Then, the untreated units are matched to the treated units according to a given matching algorithm.

We adopted the logistic regression approach to compute the propensity scores. Given a set ${\bf X}$ of measured covariates (i.e., galaxy properties), 
the coefficients $\bm{\beta}$ are linearly fit according to the linking function defined as
\begin{equation}
 {\rm {\bf Y}} = \ln \left( \frac{\bm{\mu}}{1 - {\bm{\mu}}} \right)\ , \nonumber
\end{equation}
\noindent where ${Y}_i = {X}_i \beta_i$, and the propensity scores are then given by
\begin{equation}
 \bm{\mu} = \frac{1}{1 + \exp(- {\rm {\bf Y}})}\ . \nonumber
\end{equation}

We used the nearest-neighbour method to perform the matching, i.e., the treated unit $i$ is matched to th untreated unit $j$ in such a way that the distance $|p_i - p_j| = \min_{k \in j}\{ |p_i - p_k| \}$. We allow control units to be discarded, and the model for distance measure is re-estimated after units are discarded. 
The match between the treated units with control units are made in random
order\footnote{Other two options are matching from the largest value of the
  distance measure to the smallest and the other way around.}, and we
perfomed the matching procedure $1000$ times. In each time, we create a
control sample with twice as many objects as the treated sample, and computed the KS $p$-values for all the galaxy properties. The values shown in Figs.~\ref{Fig_PSM} to \ref{Fig_Rrvir_sbgg} correspond to the median, $1$, and $99$-percentiles of the resulting distributions of $p$-values.

We first performed the match by \MvirNoUnit\ and \MstarNoUnit, as shown in Fig.~\ref{Fig_PSM_ap}. Around $48\%$ of the galaxies in the untreated control sample reside in haloes with \Mvir~$> 13.7$ (Fig.~\ref{Fig_PSM_ap}a). On the other hand, only $10\%$ of the groups with \deltam~$> 2.0$ are more massive than \Mvir~$> 13.7$. As a consequence, untreated control units with \Mvir~$< 13.7$ are very likely be matched to the treated sample, regardless of their \MstarNoUnit\ value (i.e., the coefficient $\beta$ for \MvirNoUnit\ is much greater than that for \MstarNoUnit), and the stellar mass becomes less important during the matching procedure. Since many studies suggest that the galaxy properties are more correlated with stellar mass than to the environment where the galaxy reside (e.g. \citealp{Balogh.etal:2009,
McGee.etal:2011, Trevisan.etal:2012, Woo.etal:2013}, but see also \citealp{Peng.etal:2010, vonderLinden.etal:2010, Mahajan.etal:2011}), it is desirable to get an agreement between the stellar mass distribution better than the one shown in Fig.~\ref{Fig_PSM_ap}b.

To overcome this issue, one option would be using the propensity score as only part of the matching distance, adding another distance to emphasise the variable of interest (E. Cameron, private communication; see also page 8 in \citealp{Caliendo.Kopeinig:2008}). 
However, the {\sc MatchIt} package does not include variable weighting, and implementing this approach is out of the scope of this paper. 
Instead, in Section~\ref{Sec_PSM}, we restricted the comparison between SBGGs in LGGs and regular galaxies to groups with \Mvir~$<13.7$, and perfom the matching by galaxy properties only.

\bibliographystyle{mnras}

\label{lastpage}

\end{document}